\documentclass[sigconf,monochrome]{acmart}

\usepackage{subfigure}
\usepackage{verbatim}
\usepackage{multirow}
\usepackage{makecell}
\usepackage{changepage}
\usepackage{url}
\usepackage{comment}
\usepackage[group-separator={,}]{siunitx}

\newcommand{\rev}[1]{{\color{}{#1}}}

\AtBeginDocument{%
  \providecommand\BibTeX{{%
    \normalfont B\kern-0.5em{\scshape i\kern-0.25em b}\kern-0.8em\TeX}}}

\setcopyright{none}

\begin{document}

\title[\rev{A Large-scale Analysis of Patent Citations to HCI Research}]{\rev{Breaking Out of the Ivory Tower: \\A Large-scale Analysis of Patent Citations to HCI Research}}

\author{Hancheng Cao}
\affiliation{
  \department{Computer Science}
  \institution{Stanford University}
  \city{California}
  \country{United States}}
\email{hanchcao@stanford.edu}

\author{Yujie Lu}
\affiliation{
  \department{Computer Science}
  \institution{University of California, Santa Barbara}
  \city{California}
  \country{United States}}
\email{yujielu@ucsb.edu}

\author{Yuting Deng}
\affiliation{
  \department{School of Computer Science}
  \institution{Carnegie Mellon University}
  \city{Pennsylvania}
  \country{United States}}
\email{yutingde@andrew.cmu.edu}

\author{Daniel A. McFarland}
\affiliation{
  \department{School of Education}
  \institution{Stanford University}
  \city{California}
  \country{United States}}
\email{dmcfarla@stanford.edu}

\author{Michael S. Bernstein}
\authornote{Corresponding author.}
\affiliation{
  \department{Computer Science}
  \institution{Stanford University}
  \city{California}
  \country{United States}}
\email{msb@cs.stanford.edu}


\renewcommand{\shortauthors}{Hancheng Cao et al.}
\begin{abstract}
What is the impact of human-computer interaction research on industry? \rev{While it is impossible to track all research impact pathways, the growing literature on translational research impact measurement offers patent citations as one measure of how industry recognizes and draws on research in its inventions.
In this paper, we perform a large-scale measurement study primarily of \num[group-separator={,}]{70000} patent citations to premier HCI research venues, tracing how HCI research are cited in United States patents over the last 30 years. We observe that 20.1\% of papers from these venues, including 60--80\% of papers at UIST and 13\% of papers in a broader dataset of SIGCHI-sponsored venues overall, are cited by patents---far greater than premier venues in science overall (9.7\%) and NLP (11\%).} However, the time lag between a patent and its paper citations is long (10.5 years) and getting longer, suggesting that HCI research and practice may not be efficiently connected. 

\end{abstract}



\maketitle

\section{Introduction}


What is the impact of human-computer interaction research beyond academia? Does HCI research diffuse into the industry\footnote{\rev{In this paper, we use `industry' to refer to non-research efforts that aim at practical impacts, e.g. patents, products, design practices, which usually target a broad audience than academic researchers. Thus, in this paper, industry labs whose primary focus is to publish research papers are considered academia rather than industry.}}, contributing to technological inventions and products? \rev{Are most its insights ignored by the industry?} As an applied field of study intended to be closely relevant to application --- where a considerable proportion of our research community's contributions are functional prototypes and design implications for practitioners --- the answers to these questions are critical to evaluating our translational success. There have been rich discussions regarding the industry impact of HCI research since the early years of the field, and the relationship between research and practice in HCI has long been a focal subject in both research papers~\cite{colusso2017translational, colusso2019translational} and conference panels~\cite{czerwinski2003hci,buie2010bring, chilana2015technology}.

\rev{The literature remains unclear on the field's level of success in achieving this impact. One line of the literature suggests high barriers: that HCI research has remained distant from industry impact, and that ``HCI researchers and HCI practitioners work in relatively separate spheres of influence''~\cite{czerwinski2003hci}. This line of work also argues there is a considerable research-practice gap, one that is ``real and frustrating''~\cite{norman2010research} and likely the result of a long list of barriers~\cite{colusso2017translational, winkler2013research}. However, another line of literature argues that the field achieves considerable success, that ``HCI is at the vanguard of innovation and has repeatedly influenced industry''~\cite{harrison2018hci} and that ``there is no question that research in the area of user interface software tools has had an enormous impact on the current practice of software development''~\cite{myers2000past}.}

\rev{These threads of work are not necessarily incompatible---high barriers do not rule out the existence of successes that overcome these barriers---but the field's overall status remains unclear: how far have we come, and how far do we have to go?} One approach toward resolving this debate is to pursue new methods for measuring HCI's impact. Prior work has developed rich in-depth qualitative evidence ranging from personal technology transfer experience~\cite{czerwinski2003hci} to interviews with multiple stakeholders involved in the translation process~\cite{chilana2015user}. Yet as the HCI community grows and both well-known successes and painful failures become easier to point to, it becomes more and more urgent that we also assess broader patterns.

To fill this gap, we draw on methods from the growing measurement literature on innovation in translational sciences~\cite{ahmadpoor2017dual, breschi2010tracing, li2017applied, manjunath2021comprehensive, yin2022public}, where patent citations to research have been regarded as a valuable proxy of the impact that science has on industrial practice. \rev{While patent citation to research citation does not directly guarantee industry impact, it reveals one potential pathway through which industrial inventors are aware of and recognize research articles: a necessary but not sufficient step towards industry impact.\footnote{\rev{More discussion and reflection on the usage of patent citation to science to study industry impact of research in Section~\ref{method} and Section~\ref{limitation}}}} Work using this approach has revealed the relevance of research and practice across science~\cite{ahmadpoor2017dual}, mapped the translation landscape in bio-medicine~\cite{li2017applied,  manjunath2021comprehensive}, and demonstrated that referencing science in the invention is associated with greater practical value~\cite{hicks2000research}. 

Leveraging the modern analysis approaches from this line of work~\cite{marx2020reliance, marx2022reliance}, we report the first large-scale quantitative analysis of how HCI research is (and is not) being cited by patents. \rev{In doing so, we focus on one possible route of industry impact through HCI research: patents. There are many types of contributions in HCI---design patterns, behavioral results and theory among many others---and a patent lens focuses us only on styles of contribution that are considered prior art for patents, often systems and interaction contributions. }Specifically, we draw on data from Microsoft Academic Graph, Semantic Scholar, the United States Patent and Trademark Office (USPTO), and linkages between them~\cite{marx2020reliance, marx2022reliance}. This dataset enables us to study research papers from four premier venues in HCI, including CHI, CSCW, UIST, and UbiComp, \rev{and then replicate across all 20 SIGCHI sponsored venues that appear in Microsoft Academic Graph}, tracing how those research papers are cited in patent documents from the 1980s through 2018. We study the institutes involved in the process, leverage citation analysis to measure the number and proportion of papers cited by patents over time and measure the length of time it takes before a paper is recognized by patents. We further conduct textual analysis to understand the topics that are likely to be cited in patents, and compare how patent-cited research differs from its non-patent cited counterparts. 

We observe that: (1) \rev{HCI research has been cited extensively by patents --- overall 20\% of papers from CHI, CSCW, UIST and UbiComp, and 13.4\% of SIGCHI sponsored venues, are patent-cited, including a surprising 60-80\% of UIST papers over a twenty year period, higher than  1.5\% of science overall and 7.7\% of biomedicine;} (2) The patent-paper time lag is long (on average 10.5 years) and is getting longer, such that citations from academic HCI research have dropped off by the time a paper receives patent attention; and (3) Within HCI research, there is substantial heterogeneity in patent citations across topics, for example, interaction and input techniques research are especially likely to be referenced by patents while theory, social and experience design research are not. This analysis provides the first quantitative survey of the HCI technology transfer landscape. While acknowledging potential limitations of patent citation as a method, we conclude that HCI has had a considerable impact on industry and is finding more relevance to practice than most disciplines in science. Yet, it takes a long time for innovations in academia to be recognized and taken up by industry, corroborating the ``long nose'' theory on HCI innovation~\cite{buxton2008long, harrison2018hci}. 



The contributions of this paper are as follows: 
\begin{itemize}
\item{We introduce measuring patent citations to science as a novel method to study research-practice relationships in HCI. This provides quantitative evidence that complements qualitative evidence in existing HCI literature. We release our analyzed dataset to enable future analysis.\footnote{Available dataset at: \url{https://doi.org/10.7910/DVN/QM8S1G}.}}
\item{We present the first large-scale, empirical study measuring the translational, longitudinal landscape of HCI research from paper to patent inventions with comparisons to other fields. This allows us to  better understand and evaluate how HCI as an applied field is or is not finding connections to practice.} 
\item{Our work contributes to reflections and recommendations for the HCI community to better foster a translational environment and recognize impacts beyond academia.}

\end{itemize}

\section{Background and Related Work}
In this section, we position our work in the literature on \rev{industry impact}, the HCI research-practice divide, and bibliometric analysis in HCI. 

\subsection{\rev{Industry impact}}
\rev{Industry impact are often achieved through technology transfer, which refers to the transmission of knowledge generated by an individual, the university, government agencies, or any institution capable of generating knowledge, to another person or organization, in an attempt to transform inventions and scientific outcomes into new products and services that benefit society~\cite{mendoza2018systematic}.} Government and funding agencies (e.g., in the United States, NSF and NIH) increasingly seek to nurture ``translational research'' to facilitate \rev{industry impact} from basic research so as to generate greater applied value and promote technology advances~\cite{woolf2008meaning,zerhouni2003nih}, and prior research has shown inventions that refer to high-quality research are more likely to be great inventions of value~\cite{poege2019science, hicks2000research}.

Prior research has sought to identify when, where, and how scientific research influences industry invention~\cite{backer1991knowledge, breschi2010tracing, cleary2018contribution, li2017applied}. \rev{There, patent citations to science have been widely used as a proxy for studying technology transfer from research to practice despite noises, as it is one of the only available large-scale records of the knowledge flow from research to practice that demonstrate the initial awareness and recognition of research in industrial inventions.} For instance, ~\citet{tijssen2001global} revealed through patent-paper citations how Dutch-authored research papers influence inventions. Likewise, ~\citet{ahmadpoor2017dual} studied 4.8 million US patents and how they link to 32 million research articles, finding that over half of patents cite back to a research article and that patents and papers are on average 2--4 degrees separated from the other domain, providing some insight into the interplay between patents and prior research. ~\citet{jefferson2018mapping, manjunath2021comprehensive} used patent citations to science data, measuring and reporting statistics describing how research in biomedicine turns into inventions. ~\citet{liaw2014can} proposed a method to rank academic journals that utilizes non-patent references in patent documents to evaluate their practical impact. Other works used patent citation to science to study the strategy of inventors (e.g. deep search vs. wider scope search) and how the strategy relates to technology impacts and organization performance~\cite{katila2002something, gittelman2003does, fleming2004science,arts2018paradise}. To facilitate further studies on how inventions rely on basic science, ~\citet{marx2020reliance,marx2022reliance} linked and disambiguated patent citations to science linking the USPTO dataset and Microsoft Academic Graph.\footnote{We leverage this particular dataset in our analysis.}

\rev{We build off this rich social science literature by studying industry impacts of HCI research through leveraging and extending their methods\cite{harrison2018hci}.}

\subsection{From HCI research to practice}
HCI is a field that emphasizes the design and the use of computer technology, especially interfaces between people and computers. HCI research implement, demonstrate and test new technologies through prototyping and end-user feedback~\cite{lindley2017implications}, and most HCI work includes `design implications' sections aiming to translate their research insights to more practical outcomes. The applied nature of HCI lead to the community's long-standing interest in \rev{industry impact}, with many publications and panel discussions at conferences aimed at facilitating better technology transfer~\cite{czerwinski2003hci,chilana2015technology, kawas2021translational}. \rev{One line of the literature primarily focus on the many barriers HCI faced in translating research insights to industrial practice~\cite{colusso2017translational,czerwinski2003hci}, while another line of literature speaks to the considerable impact that HCI research has had or could have on the industry~\cite{harrison2018hci,myers2000past,shneiderman2016new}.}

Many papers argue that despite the insights that HCI research can offer to practitioners, HCI research findings are rarely used in industry~\cite{colusso2017translational}: that there has been an ``immense'' research practice gap in practice that is ``real and frustrating''~\cite{norman2010research}, that ``HCI researchers and HCI practitioners work in relatively separate spheres of influence''~\cite{czerwinski2003hci}, and that ``attendees at venues like ACM CHI often lament that no HCI research ever goes into product''~\cite{harrison2018hci}. ~\citet{colusso2017translational} interviewed design practitioners so as to understand why they do not use academic research and why and how they use other resources in their works, presenting a detailed catalog of barriers that inhibit academic resources usage in industry, such as the content being hard to read, hard to find, and not actionable. ~\citet{chilana2015user} stated the distinct goals of HCI research and product may result in a research-practice gap, that the users who are the major focus of the user-centered design approach in HCI research are generally not the buyers of HCI products, and that to make a research-to-product transition one has to switch from being user-centered to adoption-centered. Furthermore, prior work~\cite{czerwinski2003hci, winkler2013research} suggested that HCI researchers usually lack the knowledge, resources, connections, experience, interest, or time to pursue technology transfer. Other work has shown similar results demonstrating a research practice gap in HCI~\cite{buie2013practice, geldof2007practitioner}. 

Prior research has discussed potential approaches to address the research-practice gap. For instance, ~\citet{velt2020translations} identified two key dimensions of the research-practice gap -- general theory vs. particular artifacts, and academic HCI research vs. professional UX design practice -- and discussed the benefits of translation led by researchers, by practitioners, or co-produced by both as boundary objects. 
~\citet{colusso2019translational} proposed a continuum translational science model for HCI that consists of three steps: basic research, applied research, and design practice. 
\rev{\citet{shneiderman2016new} wrote a book proposing principles to better blend science, engineering and design to achieve innovations and breakthroughs.}
Other work discusses the challenges and lessons learned from the specific translation of HCI research to practice~\cite{remy2015bridging, rynes2012research}.

\rev{Meanwhile, another line of work argues that HCI research could have considerable impact on industrial practice despite the barriers}. Harrison argues that ``HCI is at the vanguard of innovation and has repeatedly influenced industry [...] HCI research has a much greater impact in identifying opportunities in the first place, establishing the science and methods, building a shared vision, and developing a pipeline of human talent''~\cite{harrison2018hci}. Likewise, ~\citet{myers2000past} wrote ``There is no question that research in the area of user interface software tools has had an enormous impact on the current practice of software development. Virtually all applications today are built using window managers, toolkits, and interface builders that have their roots in the research of the 70’s, 80’s, and 90’s''. Shneiderman's work ~\cite{shneiderman2019growth} further stated that ``The remarkably
rapid dissemination of HCI research has brought profound changes that enrich people’s lives'', but also providing a tire-tracks diagram showing how HCI research on subjects such as hypertext, direct manipulation, etc. turned into product innovations by industry. Similarly, product innovations over the years mirror the early ideas of canon HCI visions~\cite{weiser1999computer, bush1945we}. Other research detailed successful cases of tech transfer, such as the translation of the multi-touch interface from research into the Apple iPhone and Microsoft Surface, while highlighting a long time lag between initial research and commercialization, which can be 20 years or more~\cite{buxton2008long, harrison2018hci, shneiderman2019growth}.

This prior work guides us to the following research questions:

\textbf{RQ1}: \textit{What} is the impact of HCI research on patents? How much HCI research is cited in patents?

\textbf{RQ2}: \textit{When} is the impact of HCI research on patents? How long does that impact take?

\textbf{RQ3}: \textit{Where} is the impact of HCI research on patents? Which topics of research are especially likely or unlikely to diffuse?

\textbf{RQ4}: \textit{Who} is involved in the process of recognizing HCI research on patents? Which institutions produce such work, and which consume it?






The rich qualitative insights derived from case studies, fieldwork, interviews, and personal experience, open an opportunity for complementary work that engages in quantitative, longitudinal analysis that directly measures how HCI research gets recognized in industry inventions and technologies. We believe that such a viewpoint might systematically detail the translation landscape of HCI as a field. 

\subsection{Bibliometrics and HCI}
As an important area of computing and information science, HCI has featured several projects (e.g.,~\cite{kaye2009some, mack2021we}) that quantitatively understand the structure and evolution of the field through the study of writing and citation patterns, known as bibliometrics~\cite{fortunato2018science}.

One commonly used bibliometric method is an analysis of a large-scale citation network, which leverages the increasingly available citation data from publishers such as Web of Science and Microsoft Academic Graph and their associated metadata of the scientific publications (e.g. institutes, authors), and even textual analysis (e.g. topic modeling, keyword extraction) of the scientific publications, so as to gain insights on patterns behind the diffusion of scientific ideas~\cite{fortunato2018science, wang2013quantifying}, research productivity~\cite{liu2018hot, wang2019early}, and identify potential ethical and social issues in science~\cite{hofstra2020diversity, kim2022gendered}. For instance, ~\citet{koumaditis2017human} leveraged citation data from 962 HCI publications and reveal that HCI research can be categorized into major themes of design, data management, user interaction, psychology, and cognition, and they identified more recent trends in HCI in the workplace, sensors, and wearables. Likewise, ~\citet{kaye2009some} reported ``some statistical analyses of CHI'', including author counts, gender analysis, and representations of repeat authors so as to motivate discussions on the preferred state of CHI. ~\citet{bartneck2009scientometric} reveal that only a small percent of countries account for the majority of CHI proceedings, and present a ranking of countries and organizations based on their H-index of CHI proceedings. ~\citet{correia2018scientometric} used 1713 CSCW publications and characterized top CSCW papers, citation patterns, prominent institutes as well as frequent topics, highlighting the fact that CSCW is influenced primarily by a few highly recognized scientists and papers. The authors further quantitatively explored the relationship between collaboration types and citations, paper frequency, etc~\cite{correia2019effect}. Similar types of analysis have also been done on more regional HCI conferences~\cite{gupta2015five, nichols2015scientometric, mubin2017scientometric,bartneck2011end} as well as studying subcommunities in HCI~\cite{mack2021we, wang2021bibliometric, wania2006design}. 

Visual analytics is another approach used to help understand HCI's evolution. For instance, ~\citet{lee2005understanding} proposed a system PaperLens to reveal trends, connections, and activity of 23 years of the CHI conference proceedings. ~\citet{matejka2012citeology} proposed an interactive visualization that highlights family trees of CHI and UIST papers. ~\citet{henry200720} presented a visual exploration of four HCI conferences. They showed that the years when a given conference was most selective are not correlated with those that produced its most highly referenced articles and that influential authors have distinct patterns of collaboration.

To the best of our knowledge, there have been no analyses leveraging quantitative methods to study recognition of HCI research beyond academia as we present in this article. In contrast with prior work, we leverage large-scale patent citations to quantify the impact of HCI research in practice.



\section{Method}
In this section, we describe the method we used to study the impact of HCI research papers in practice using patent citations to science. 

\subsection{\rev{Patent citations as a pathway to study industry impact of research papers}}\label{method}

We leverage patent citations to research as a proxy to study the influence of HCI research on industrial practice at scale. \rev{While patent citation to research citation does not directly mean industry impact, it reveals one important potential pathway from research to practice where industrial inventions become aware of and recognize research articles, which is often a necessary but not sufficient step towards producing industry impact. Alongside with studying other forms of influence, such as design processes (e.g., usability testing, heuristic evaluation), design patterns, open source software (e.g., d3, Vega), patent citations to science could help us piece together the translational landscape in HCI}. This method is widely used in the innovation literature (e.g.,~\cite{ahmadpoor2017dual,manjunath2021comprehensive,marx2020reliance,katila2002something, gittelman2003does, fleming2004science}). Patent citations to research are considered valuable signals indicating the influence of research on the industry, signals that ``reflect genuine links between science and technology.”~\cite{tijssen2001global}, and ``appear to be a substantive if a noisy indicator of the role of specific, prior scientific advances''~\cite{ahmadpoor2017dual}. While citations between research articles capture research influence~\cite{fortunato2018science}, patent-to-research citations capture ``how basic research influences commercialization and thus provides a complementary measure of impact''~\cite{manjunath2021comprehensive}. Such data has been used extensively to measure knowledge spillovers from academia and government to industry~\cite{ahmadpoor2017dual,de2020citation, marx2020reliance}. 

The rationale behind the validity of this approach is that in patented inventions, inventors are obliged to disclose \textit{any} ``prior art'' related to their invention, i.e., all information known to that individual to be material to patentability”,\footnote{https://www.uspto.gov/web/offices/pac/mpep/mpep-2000.pdf} including  materials that the inventors leveraged in the invention process, or other similar material to the focal invention in order to distinguish it. The prior art includes both references to prior patents, and references to non-patent literature, such as academic articles. Patent citation is an important part of a patent, as missing prior art (either prior patents or non-patent literature), could have potential legal issues. Apart from citations provided by inventors, patent examiners who review patents for approval or rejections also add references they think are of relevance to ensure the legitimacy of the patent. 

Prior work has validated this method. Nagaoka et al.~\cite{nagaoka2015use} surveyed 843 inventors finding patent citations to science are indeed important linkages to science, despite possible errors of over- and under-inclusion. Callaert et al.~\cite{callaert2014sources} interviewed 36 inventors and report 44\% of patent citations to science are considered as ``important'' or ``very important'', and another 34\% are ``background'' citations. Based on the rich literature in this space, we conclude that patent citation to science can be used as a reliable data source to measure the recognition of HCI research efforts in inventions, thus providing a valuable proxy of HCI research impact in the industry. \rev{Of course, there is no perfect appoach for studying industry impact: we discuss and reflect on the limitations of our method in detail in Section~\ref{limitation}, and it is especially important to bear in mind there are multiple translational gaps in HCI research \cite{colusso2019translational}, and we are only studying one important step in the process with regard to patent, where certain types of contribution such as theory are likely to be under evaluated through this dimension.}

\begin{figure*}
  \centering
      \begin{subfigure}[Patent US8239784 frontpage with abstract, inventors, assignee etc.]{
        \label{fig:Email_Day}
        \includegraphics[width=0.8\textwidth]{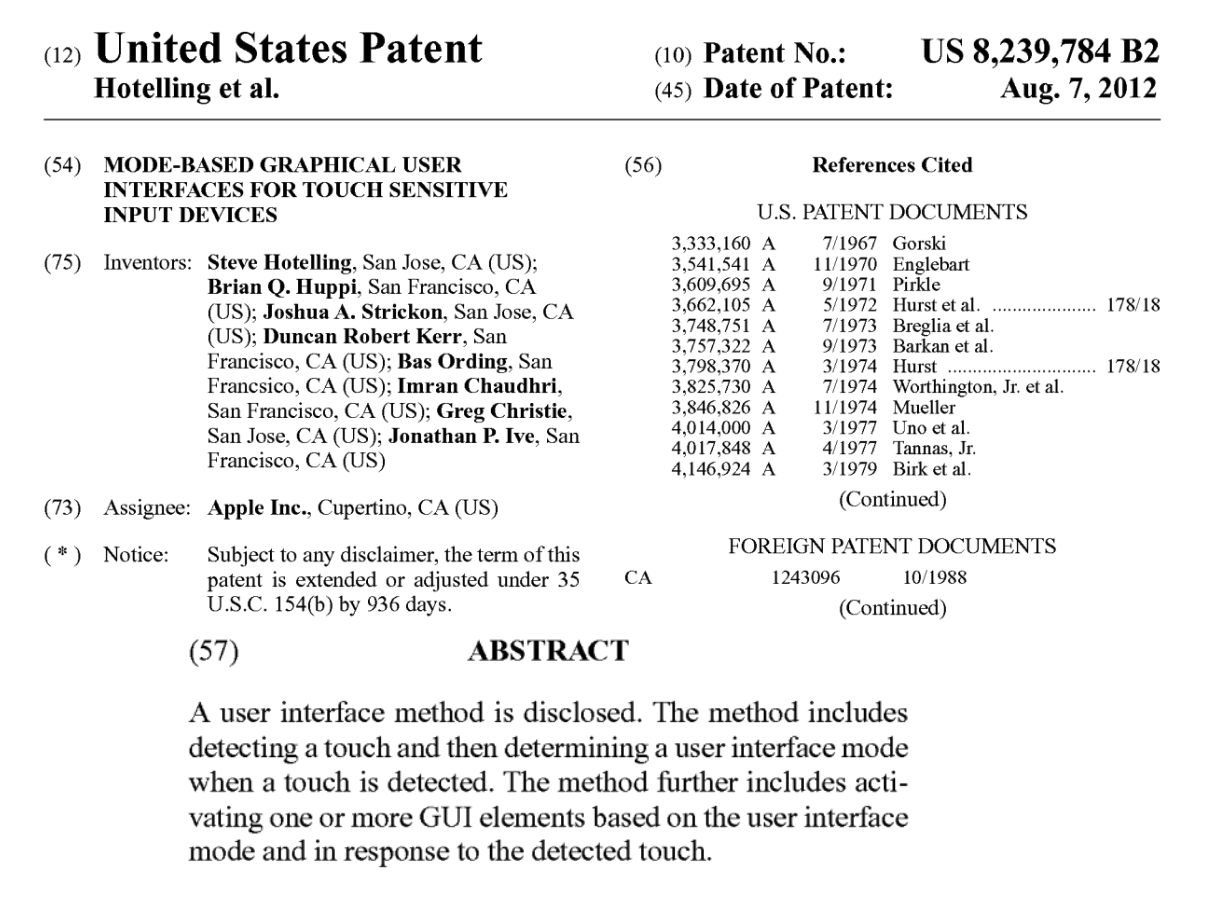}}
       \end{subfigure}
       \begin{subfigure}[Part of the citation list of Patent US8239784.]{
        \label{fig:Meeting_Day}
        \includegraphics[width=0.4\textwidth]{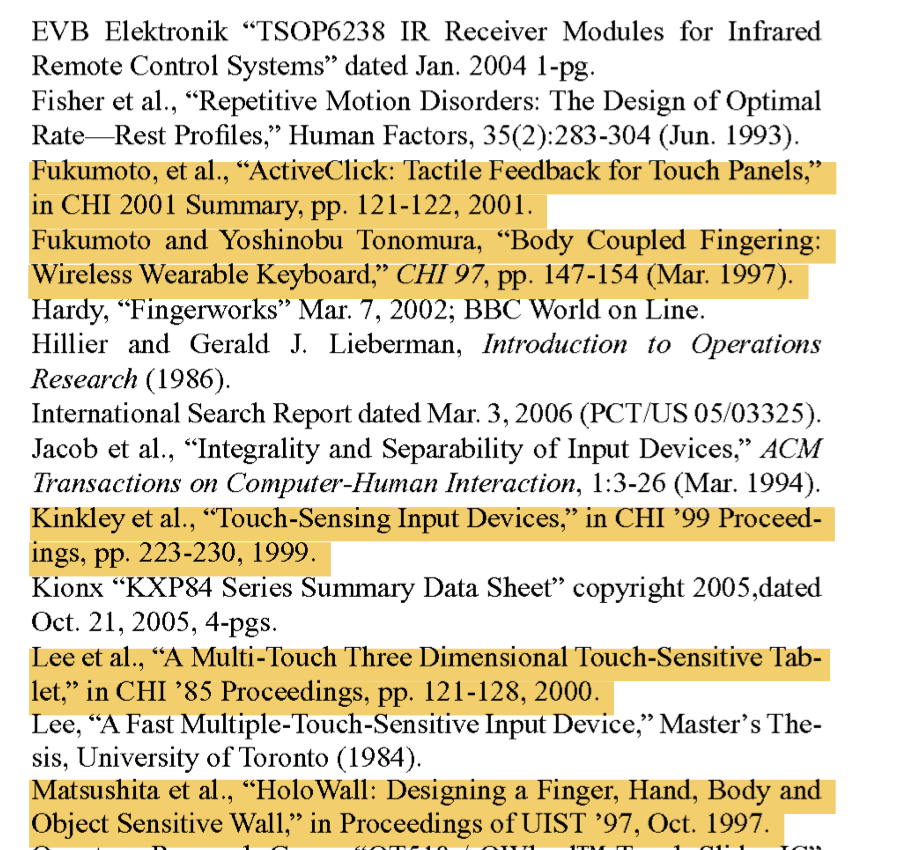}}
       \end{subfigure}
  \caption{Patents are obliged to cite prior art, including prior patents and non-patent literature (e.g. research articles). Here, a patent by Apple Inc., ``Mode-based Graphical User Interfaces for Touch Sensitive Input Devices''~\cite{hotelling2012mode}, has citation to relevant HCI papers, including ``ActiveClick: Tactile Feedback for Touch Panels'', ``A Multi-Touch Three Dimensional Touch-sensitive Tablet'', a mis-named citation to Ken Hinckley (``Kinkley et al.''), and many other references to HCI research.}
  \label{fig:patent_screenshot}
\end{figure*}

\rev{Empirically, we find support for the validity of using patent citations to research as a proxy of impact in industry. We manually check patent reference lists of a number of patents. As shown in Figure \ref{fig:patent_screenshot}, the highly-cited patent by Apple Inc. ``Mode-based graphical user interfaces for touch sensitive input devices'' (cited 1,898 times),\footnote{\url{https://patents.google.com/patent/US8239784B2/en}} cites closely related research papers in CHI on multi-touch, such as ``A Multi-Touch Three Dimensional Touch-sensitive Tablet", which is the case of technology transfer discussed by Buxton~\cite{buxton2008long}. The even more well-cited Apple Inc. patent (cited 4,018 times) ``Method and apparatus for integrating manual input'' \footnote{\url{https://patents.google.com/patent/US6323846B1/en}} also made reference to several relevant HCI papers. These cases motivate us to leverage patent citations as a signal indicating the invention's recognition of research.}



\subsection{Dataset}
\label{sec:dataset}
\rev{To study how HCI papers are recognized by patents, we required a citation graph from patent to research, and the metadata (e.g., author name, affiliation, publication year, title, venue) from both the paper side and patent side.}
The data preparation pipeline is composed of three steps: 1)~Prepare metadata of papers and patents, and the citation graph from patents to research, 2)~Select papers from the venues of interest and clean the data, and 3)~Link the clean metadata based on the citation graph.
This pipeline could be applied to other research communities, or other venues within SIGCHI, by selecting other venues of interest.

\paragraph{\textbf{Patent citation to science that connects USPTO to Microsoft Academic Graph}}
To capture references from patents to HCI research papers, we drew on a public dataset~\cite{marx2022reliance, marx2020reliance2}. This dataset is a state-of-the-art approach to connect each patent reference in USPTO (1947-2020) to academic papers (1800-2020) from Microsoft Academic Graph through matching unstructured front-page and in-text references in patents to published papers using a disambiguation matching method, resulting in $22$ million patent citations to research papers (known as Patent Citation Science dataset).\footnote{Specifically, we used the patent‐to‐article citations of Version v$37$ (Jul 19, 2022) at Zenodo: \url{http://relianceonscience.org}} In their papers, the dataset creators verified the quality of their datasets through manual checking and error analysis.
We captured the reference type (e.g., from applicant, from examiner, unknown), \rev{whether the reference appears in-text or on front page}, the time between paper publication and the citing patent application, and whether a patent citation is a self-citation to a research paper by one of the patent authors.
A paper to patent pair is considered self-cited when there is an overlap between the inventors of the patent and the authors of the cited scientific papers.

\paragraph{\textbf{Microsoft Academic Graph Metadata}}
The Microsoft Academic Graph is a heterogeneous graph that provides scientific publication records, citation relationships, the information of authors, institutions, journals, conferences, and fields of study.
We leveraged the public Microsoft Academic Graph dataset provided at Zenodo Reliance on Science project site\footnote{\url{http://relianceonscience.org}} so as to extract information with regard to academic publications, e.g., title, author, author affiliation, and year.

\paragraph{\textbf{USPTO metadata}}
We leveraged US patent data from the United States Patent and Trademark Office (USPTO)\footnote{\url{https://patentsview.org/download/data-download-tables}} to represent technological inventions. Patents have similar fields as academic publications, e.g., title, abstract, inventor,  assignee, and year. 

\paragraph{\textbf{Semantic Scholar (abstract, citation)}}
The abstract information of the paper and their academic influence (e.g., number of published papers, citation count) are missing or hard to process in the original Microsoft Academic Graph metadata.\footnote{\url{https://docs.microsoft.com/en-us/academic-services/graph/resources-faq}}
To further expand data information about authors, papers, citations, and venues, we utilize the Semantic Scholar Academic Graph API,\footnote{\url{https://www.semanticscholar.org/product/api}} which fills in this data.

The details of the data we utilize can be found in Appendix A.

\subsection{Data Preprocessing}
\paragraph{\textbf{Venue selection.}}
In our analysis, we primarily considered four impactful Human-Computer Interaction (HCI) venues: the ACM CHI Conference on Human Factors in Computing Systems (CHI), ACM Conference On Computer-Supported Cooperative Work And Social Computing (CSCW), ACM Symposium on User Interface Software and Technology (UIST), and International Joint Conference on Pervasive and Ubiquitous Computing (UbiComp).\footnote{Starting 2017, the UbiComp conference main technical tracks consist of papers published in Proceedings of the ACM on Interactive, Mobile, Wearable and Ubiquitous Technologies (IMWUT), which we captured in our data.} \rev{For a broader footprint of HCI research, we created a second dataset of SIGCHI sponsored venues\footnote{\rev{https://sigchi.org/conferences/upcoming-conferences/}} --- a total of all 20 SIGCHI sponsored venues\footnote{\rev{Details of the venues in Appendix B}} that appear in the Microsoft Academic Graph, which covers not only large, premier venues such as CHI, but also smaller, more specialized venues such as MobileHCI and CHI PLAY. We used this second set as more representative of the overall field of HCI, to further validate our findings and compare with overall patterns reported in other fields of science in a fairer way\footnote{\rev{Note that in this paper we primarily report findings on the four chose venues rather than SIGCHI sponsored venues overall. We elect to focus on these four venues as a practical matter, as we have spent considerable manual efforts in cleaning data related to the four chosen venues to ensure data quality, as indicated in ``Data Cleaning" section, which makes our analysis more likely to reflect actual trends in these venues.}}. }

\paragraph{\textbf{Data Cleaning}}
\rev{We further conducted data cleaning on the four chosen venues by looking up papers in Semantic Scholar rather than Microsoft Academic Graph.} We found that Microsoft Academic Graph metadata sometimes wrongly classify venues such as ``Brazilian Symposium on Human Factors in Computing Systems'' as ``CHI''.
To solve this issue, we filtered out irrelevant papers by manually checking the full name of the venue column from Semantic Scholar, which proves to be of better quality.
We then applied this filtering process to all the paper and patent citations to science files by joining over the paper id.

\paragraph{\textbf{Data Linking}}
In order to better combine the paper and patent information for analysis, we linked patent data, Microsoft Academic Graph data and Semantic scholar data via the Patent Citation Science dataset.\footnote{Confusingly to HCI researchers, this is known as the ``Patent Citation Science'' (PCS) dataset. We joined information from the patent side using the field \texttt{patentid} to information from the paper side using the field \texttt{magid}.}
The joined data after 2019 has incomplete or little coverage, thus we focus our analysis on HCI research papers and patents that cite HCI papers before 2019.

\paragraph{\textbf{Final Data Statistics}}
\rev{Our final data for analysis includes 23,432 papers from the four chosen venues}, with 16,014 from CHI, 3,084 from CSCW, 1,746 from UIST, and 2,588 from UbiComp across $1980$ to $2018$.
Within these papers, we captured 69,900 citation records from patent to science, with 42,676 from CHI, 5,900 from CSCW, 17,040 from UIST, and 4,284 from UbiComp, which are associated with 30,660 patents. \rev{The broader SIGCHI sponsored venue data include 57,385 papers in total (41\% are papers from the four premier venues), 83,793 citation records (51\% are citations made to the four premier venues), and are associated with 36,024 patents in total (85\% patents cited papers from the four premier venues). }

Note that for all \rev{chosen venues}, our data includes not only main conference papers but also extended abstracts, posters and other forms of publications. We did not attempt to filter and focus our analysis only on main conference papers, given the difficulty to classify and challenge fuzzy matching based on venue name (e.g. in our dataset, many posters are not explicitly labeled as poster publications and are hard to differentiate from main conference papers).

We release our dataset at: \url{https://doi.org/10.7910/DVN/QM8S1G}.

\section{Results}



\subsection{RQ1: What is the impact of HCI research on patents?} \label{rq1}
We first study the quantity of HCI papers that are later recognized by patents and present a table of top papers cited by patents.

\begin{figure*}
    \centering
    \includegraphics[width=\linewidth]{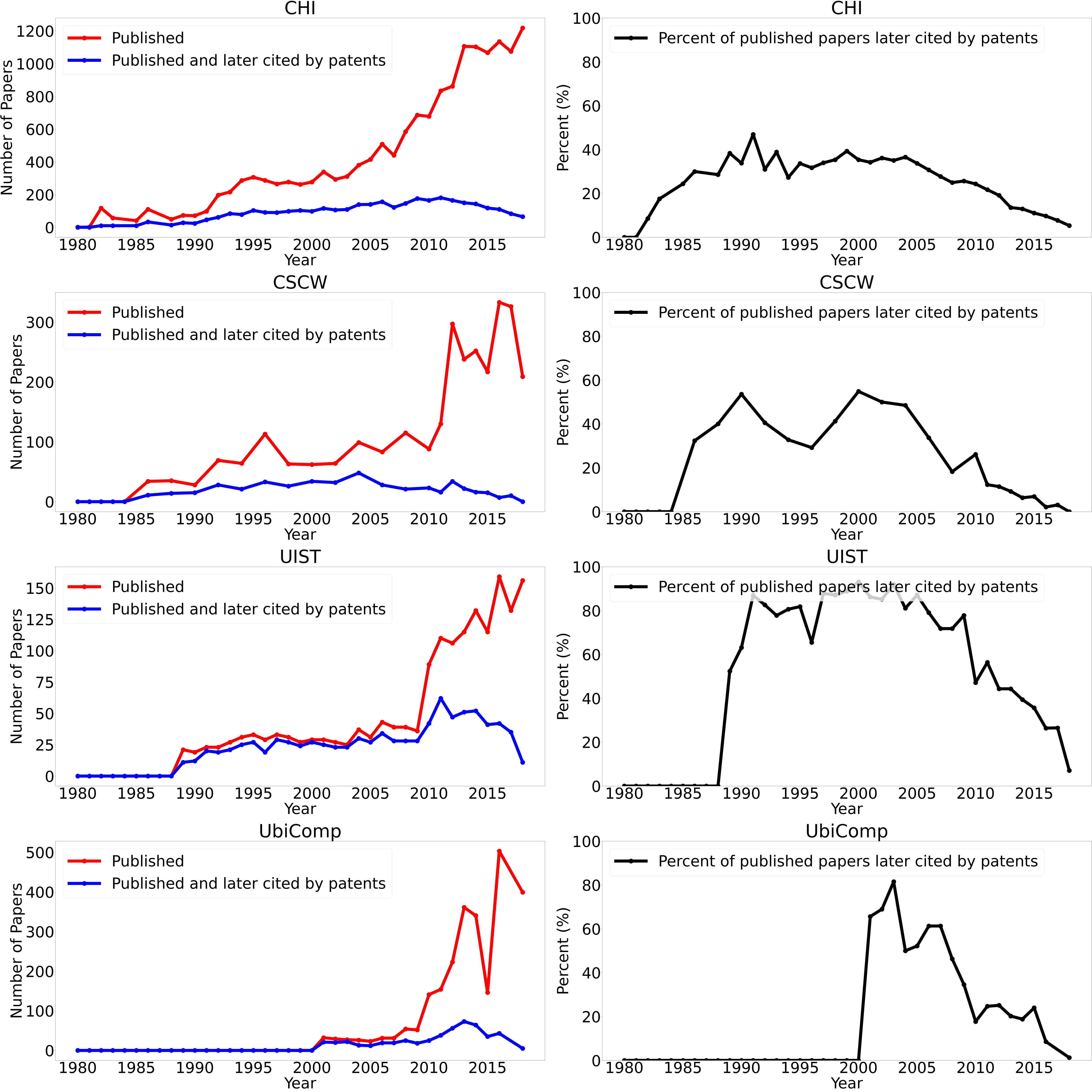}
    \caption{Left: the number of papers published by each conference per year (red) and the number of papers published in that year that were later cited by at least one patent (blue), at ACM CHI, CSCW, UbiComp, and UIST. Right: a substantial proportion of HCI papers are recognized by patents, e.g. $60\%$ - $80\%$ UIST papers are recognized by patents $1990$ - $2010$.
    } 
    \label{fig:f1}
\end{figure*}

\paragraph{\textbf{Proportion of papers that get cited by patents.}}
\rev{To assess the extent of HCI research being recognized in patents, we first calculated the aggregated proportion of the number of HCI papers at our four premier HCI venues, and SIGCHI sponsored venues overall, that were cited by patents. We found $20.1\%$ of papers in the four venues, and $13.4\%$ of papers from SIGCHI sponsored venues overall, are recognized by patents. This rate is much higher than the proportion of science cited by patents overall (approximately 1.5\% \cite{marx2020reliance}), and the prominent journal paper patent rate (9.7\% across multiple scientific fields \cite{bryan2020text}). The rate is also much higher than that of bio-medicine in general, a field that has a rich tradition emphasizing translational science, which is at 7.7\% \cite{manjunath2021comprehensive}. We replicated our analysis on premier venues in other areas of Computer Science by comparing the premier HCI venue patent rate ($20.1\%$) with premier venue patent rate of other subfields, finding that AI patent rates (as measured through AAAI and IJCAI, two of the largest and premier AI conferences) are ~5\%, Natural Language Processing patent rates (as measured through ACL, EMNLP, and NAACL, three of the largest and premier NLP conferences) are ~11\%, and Computer Vision patent rates (as measured through CVPR, ECCV, and ICCV, three of the largest and premier computer vision conferences) are ~25\%. Two-proportion z tests further confirm the significance of the difference in percentages with $z$ = 51.1, 23.9, -13.1, ~$(p<.001)$ when comparing premier HCI venue patent rate with patent rates of premier venues in AI, Natural Language Processing and Computer Vision. 
Taken together, these results suggest that HCI's impact through patent citations is higher than science overall, biomedicine, AI, and NLP, and roughly at par with Computer Vision, an area of intense industry interest. }


\rev{Are research citations in patents truly central to the patents, or are they thrown in just to satistfy a patent examiner? To answer this question, we leverage a distinction between in-text citations and front page citations in patents. This distinction allows us to more directly measure the impact of HCI research in patents. In-text patent citation to science, as suggested by prior work~\cite{marx2022reliance,bryan2020text}, are more likely to ``capture the scientific articles upon which the scientists truly relied upon for inspiration'' and ``have the potential to more accurately represent the sources of scientific inspiration upon which the inventors actually drew in the invention process" since they ``tend to be supplied by the inventors themselves'', in contrast to ``legally binding'' front page citations which ``tend to be carefully reviewed (and sometimes added) by patent attorneys''~\cite{marx2022reliance}. We find 4.1\% papers in our chosen four venues have been cited in-text by patents, whereas the proportion of patent in-text citation to science is 2.3\% for SIGCHI sponsored venues and 1.4\% for science overall. This result further replicates our finding that HCI research appears to have real impact, surprisingly even moreso than many other fields.}


\rev{Investigating temporal patterns, we plot the total number of HCI research papers in each of the four venue published over years, shown in red in Figure~\ref{fig:f1}.} HCI research has grown rapidly over the past $38$ years for all four venues, especially at CHI: from $74$ papers in $1982$ to ~\num{1200} in $2018$. This growth is particularly pronounced within the last $10$ years. We then counted the total number of HCI papers cited by patents by the publication year of the paper and calculated the ratio between the number of HCI papers cited and the total number of HCI papers accepted in a particular year by each venue (blue line in Figure~\ref{fig:f1}).  
The citation ratios start climbing especially starting around 1990 and persist since then (Figure~\ref{fig:f1}),\footnote{We removed years where conferences did not meet from our analysis and smoothed the curve, e.g. CSCW was only held every other year until 2010.} with several conferences observing a third to a half of their papers cited by patents. At UIST in particular, the patent citation ratio reaches $60\%$ - $80\%$ from $1990$ - $2010$. 

The citation ratio decreased after 2015. 
One possible explanation is the time lag between patent and paper is long, e.g., it might take a decade for a paper to start gathering patent citations, and papers since 2015 are still too young by this metric. This time lag will be further discussed in Section~\ref{sec:patent_recognize}. In other words, the data are right censored, i.e., more recent papers have not been fully recognized by patents captured in our dataset. As such, we expect a higher proportion of HCI papers overall will be referenced by patents eventually.

\paragraph{\textbf{Increasing citations to HCI research in patents.}}

\begin{figure*}[t]
    \centering
    \includegraphics[width=\linewidth]{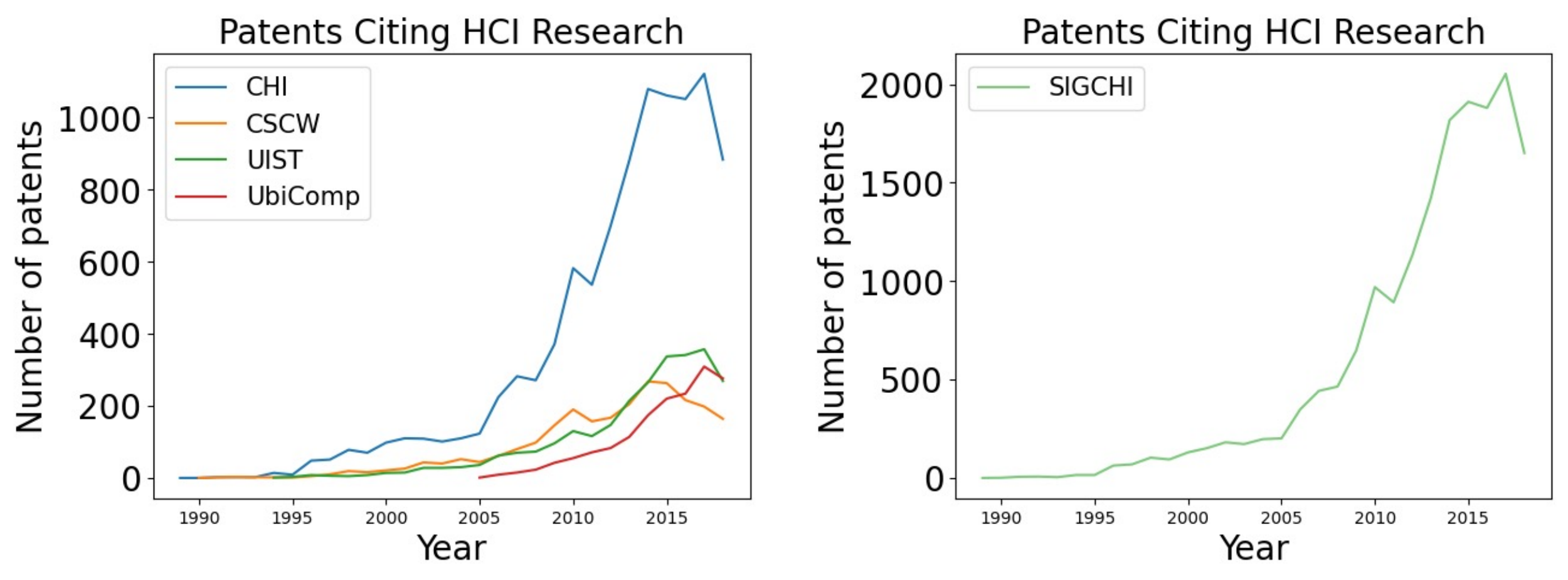}
    \caption{\rev{Left: over 1000 patents are citing CHI paper each year after 2014. The number of patents citing HCI research began rising after 2000 and more than quintupled since then. Right: the number of patents citing SIGCHI sponsored venues follow similar trend, as a large proportion (85\%) made references to the four premier venues.}
    }
    \label{fig:f2}
\end{figure*}

\rev{A total of \num{30660} patents cite research in the four chosen venue, and 36024 patents cite research from SIGCHI sponsored venues overall.} This raw volume began increasing after $2000$ (Figure~\ref{fig:f2}, and has more than quintupled since $2000$ at CHI from around 175 patents per year in 2000 to over 1000 per year in 2014). However, the number of patents plateaus and even decreases a bit in more recent years, e.g. patents begin citing less and less CSCW research starting in 2014. This could be a result of changes on the demand side, e.g., the industry is less interested in novel social computing applications, or on the supply side, e.g., HCI publishing more papers that are not intended to be as industry-relevant. More evidence is needed to derive the mechanisms behind this result, beyond the scope of our current work.

\paragraph{\textbf{Top cited papers by patents in HCI}}

We further examined the HCI papers that were cited the most by patents by each venue (Table~\ref{tab:top_HCI}). Papers highly cited by patents also tend to be highly cited by research. The papers most highly cited by patents are primarily systems work, e.g., building a new system or proposing a new design. This result parallels with the earlier observation that UIST has the highest rate of papers cited by patents since UIST is particularly targeted at new interfaces, software, and technologies. 
Most papers in this list were published prior to 2005; however, the majority of the patents that cited HCI papers come after 2005, indicating again the potential long time lag between paper publication and patent reference in Section~\ref{sec:patent_recognize}.

\begin{table*}[tb]
\centering
\resizebox{\textwidth}{!}{%
\begin{tabular}{l l l l}
\toprule
\textbf{Title} & \textbf{Patent Citations} & \textbf{Paper Citations} & \textbf{Year Published} \\
\midrule
\multicolumn{4}{c}{\textbf{CHI}} \\
\midrule
    A multi-touch three dimensional touch-sensitive tablet & 708 & 231& 1985 \\
    PaperLink: a technique for hyperlinking from real paper to electronic content & 200 & 134&1997 \\
    Bringing order to the Web: automatically categorizing search results & 196 & 486 &2000 \\
    A study in two-handed input & 175 & 544&1986 \\
    Generalized fisheye views & 175 & 2180&1986 \\
    SmartSkin: an infrastructure for freehand manipulation on interactive surfaces & 166 & 770&2002 \\
    AppLens and launchTile: two designs for one-handed thumb use on small devices & 159 & 133&2005 \\
    Active click: tactile feedback for touch panels & 156 & 195&2001 \\
    Finding others online: reputation systems for social online spaces & 153 & 100&2002 \\
    Applying electric field sensing to human-computer interfaces & 142 &272 &1995 \\
\midrule
\multicolumn{4}{c}{\textbf{CSCW}} \\
\midrule
    GroupLens: an open architecture for collaborative filtering of netnews & 185 & 5771&1994 \\
    WebSplitter: a unified XML framework for multi-device collaborative Web browsing & 166 &186 &2000 \\
    Blogging as a social activity, or, would you let 900 million people read your diary? & 121 & 584&2004 \\
    MMConf: an infrastructure for building shared multimedia applications & 106 &313 &1990 \\
    An experiment in integrated multimedia conferencing & 103 & 157&1986 \\
    Collaboration using multiple PDAs connected to a PC & 94 & 391&1998 \\
    Interaction and outeraction: instant messaging in action & 90 & 1225&2000 \\
    Providing presence cues to telephone users & 83 &177 &2000 \\
    Design of a multi-media vehicle for social browsing & 72 &331 &1988 \\
    Distributed multiparty desktop conferencing system: MERMAID & 69 &153 &1990 \\
\midrule
\multicolumn{4}{c}{\textbf{UIST}} \\
\midrule
    Sensing techniques for mobile interaction & 254 & 592&2000 \\
    The world through the computer: computer augmented interaction with real-world environments & 227 & 487&1995 \\
    HoloWall: designing a finger, hand, body, and object sensitive wall & 197 & 243&1997 \\
    A survey of design issues in spatial input & 166 & 417&1994 \\
    Tilting operations for small screen interfaces & 158 &412 &1996 \\
    Multi-finger and whole hand gestural interaction techniques for multi-user tabletop displays & 156 &527 &2003 \\
    DiamondTouch: a multi-user touch technology &153 & 1336&2001 \\
    The document lens & 135 & 416&1993 \\
    The DigitalDesk calculator: tangible manipulation on a desk top display & 132 & 324&1991 \\
    Pad++: a zooming graphical interface for exploring alternate interface physics & 131 &754 &1994 \\
\midrule
\multicolumn{4}{c}{\textbf{UbiComp}} \\
\midrule
    Validated caloric expenditure estimation using a single body-worn sensor & 113 & 83&2009 \\
    InfoScope: Link from Real World to Digital Information Space & 67 &34 &2001 \\
    Self-Mapping in 802.11 Location Systems & 63 &130 &2005 \\
    The NearMe Wireless Proximity Server & 62 & 162&2004 \\
    Predestination: Inferring Destinations from Partial Trajectories & 51 &498 &2006 \\
    UbiTable: Impromptu Face-to-Face Collaboration on Horizontal Interactive Surfaces & 40 &261 &2003 \\
    Accurate GSM Indoor Localization & 37 & 537&2005 \\
    Very Low-Cost Sensing and Communication Using Bidirectional LEDs & 34 & 157&2003 \\
    Particle Filters for Location Estimation in Ubiquitous Computing: A Case Study & 33 & 254&2004 \\
    PowerLine Positioning: A Practical Sub-Room-Level Indoor Location System for Domestic Use & 31 &152 &2006 \\

    \bottomrule
\end{tabular}
}
    \caption{Top CHI, CSCW, UIST, and UbiComp papers cited by patents. The majority of them are highly-cited papers in academia whose major contribution is a system.
    }
    \label{tab:top_HCI}
\end{table*}

\paragraph{\textbf{Highly-cited papers in academia are more likely to be recognized by patents}}

\begin{figure*}[t]
    \centering
    \includegraphics[width=.6\linewidth]{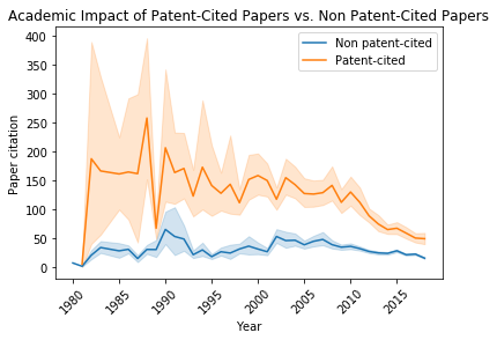}
    \caption{Papers cited by patents receive more academic citations in HCI.}
    
    \label{fig:f3}
\end{figure*}

Moreover, we investigated how academic impact relates to patent impacts, measured by the paper's number of citations from other academic papers (academic citation count) and the number of citations from patents (patent citation count). Figure~\ref{fig:f3} shows the academic citation count for both papers recognized by patents and papers not recognized by patents over time. Patent-cited papers have higher paper citations (average academic citation count 117.1)  than non-patent-cited papers (average academic citation count 27.9), a difference that is significant via an unpaired t-test~$(p<.001)$\rev{, Cohen's D=$0.58$.} 

\rev{We further conducted zero-inflated negative binomial regression\footnote{\rev{Zero-inflated negative binomial regression is ideal for modelling count-based dependent variables with zeroes, which corresponds to our data where a significant proportion of HCI papers get no patent citation.}} over patent citation and paper citation count in CHI, CSCW, UIST, and UbiComp and get regression coefficient of $0.0233$, $0.0172$, $0.0316$, and $0.0175$ respectively ~$(p<.001)$. The coefficient indicates that highly-cited papers in academia are indeed more likely to be cited by patents.} Such a relationship is especially salient at UIST.

\subsection{RQ2: When is the impact of HCI research on patents? }
\label{sec:patent_recognize}
How long does it take for patents to recognize papers? To examine this question, we investigated the time lag between patent and paper. 

\paragraph{\textbf{The time lag between patent and paper is long and getting longer.}} 
\rev{To measure how long it takes for an HCI paper to be recognized by patents, for each patent, we investigated the time lag between the issue date of the patent and the publication date of all papers it cited from our four chosen venues}. We measured the lag from the patent backward rather than from the paper forward because we cannot know whether a paper will receive a citation but has not yet---but we can know how far back a patent's citations reach.

In the four premier HCI venues, the average patent-paper lag is 10.5 years ($\sigma=6.8$ years), indicating that patents on average reference HCI research papers published 10.5 years before the patent filing date but there is significant variance over the time lag.

\begin{figure*}[t]
    \centering
    \includegraphics[width=\linewidth]{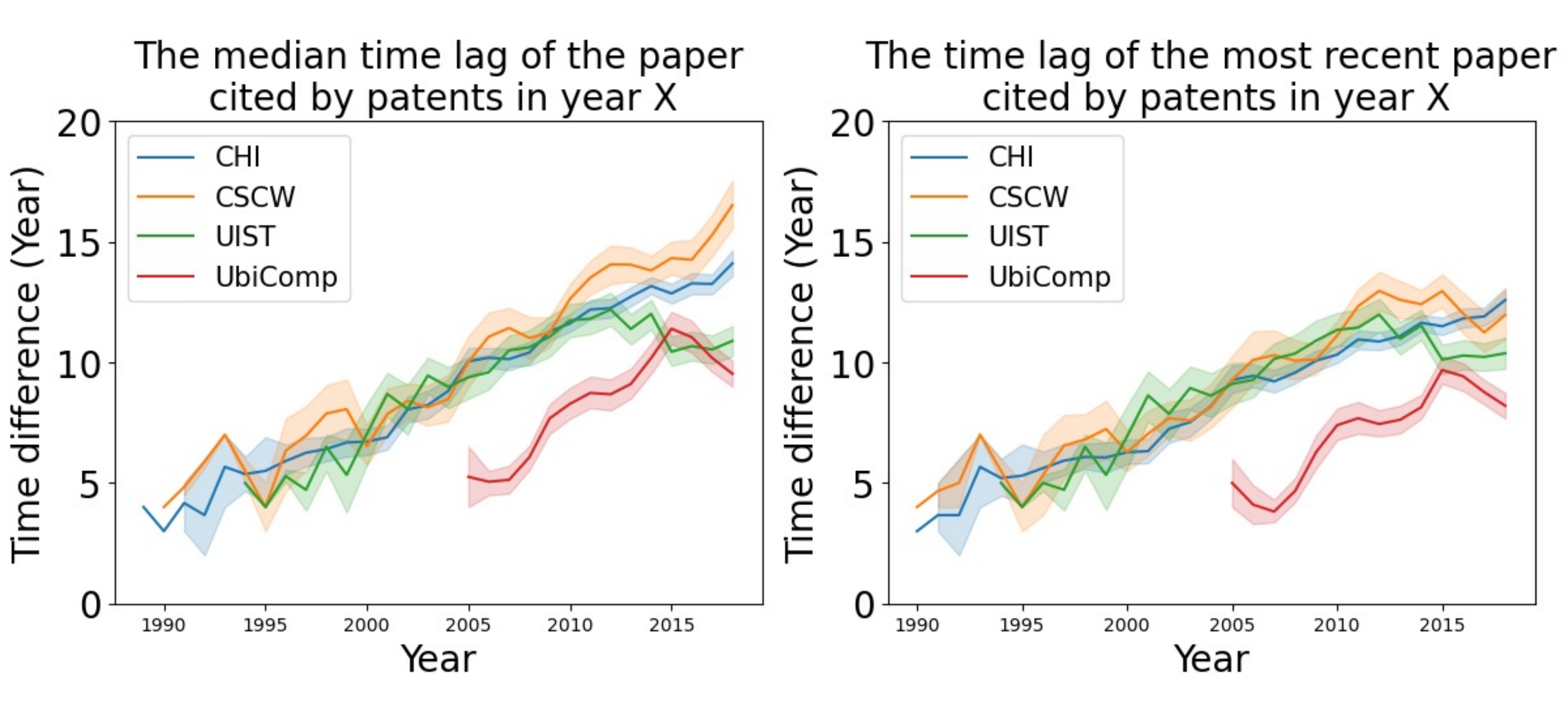}
        \caption{The time lag between patent and paper is long and getting longer across venues.}
    \label{fig:f9}
\end{figure*}

We then studied how the time lag varies over time by aggregating the patent-paper time lag at the individual patent levels. 
As Figure~\ref{fig:f9}a) shows, the median difference between the time the cited paper is published  and the time the paper is cited by the patent, is becoming larger from 1989 to 2014 for all the venues from about around $5$ years to around $10-15$ years. However, since $2014$, this trend bifurcates among different venues: the time lag for CSCW increases to over 15 years and Ubicomp decreases to about 10 years in 2017. We also noticed that all venues have nearly indistinguishable trends except Ubicomp, which has about 3 years of time lag lower than other venues. In recent years, CSCW takes the longest time to be recognized by patents, while UIST and UbiComp take a shorter time, which could be explained by the fact that more system-driven works are likely to diffuse more quickly into practice. 

We also examined the time lag between the patent and its most recent cited paper (Figure~\ref{fig:f9}b), testing how recent the freshest research is that patents draw on. These general trends are consistent with the median time lag. Again, the difference between the time its most recent cited paper was published and the time it is patented also becomes larger from $1989$ to $2011$ for all the venues, from less than $5$ years to around $10$ years. This increase gradually slowed down, leading to a slight decrease in more recent years.




The patent citation also involves different sources, some are added by the applicants/inventors, while others are added by patent examiners. The dataset we used also provides a breakdown of reference types, including applicant/inventor added, the examiner added, other, and unknown types. References added by patent examiners are generally more recent (average time lag: 6 years) than what the inventor added (average 11.8 years), although similar trends of long time lags and increasing time lags are still observed. 

All results here indicate that patents mostly cite old research, and are citing increasingly older research, which holds true across venues and reference types. This conclusion is largely identical to what is found in science in general \cite{marx2022reliance}. We replicated our analysis on other areas of Computer Science in a similar way as in Sec \ref{rq1}, finding that the time lag between patent and their referenced papers for AI, Natural Language Processing, and Computer Vision are ~17 years, ~13 years, and ~10 years respectively, suggesting similar patterns across subfields in Computer Science. 

\paragraph{\textbf{HCI research has moved on by the time a paper receives patent attention.}}

Has the HCI community left an idea behind by the time industry gets interested? Concerns circulate that HCI has a reputation for trend following and jumping to new shiny areas every few years~\cite{buxton2008long, harrison2018hci}. Are patent-cited papers still receiving academic interest by the time it starts receiving patent citations? To answer this question, \rev{for all papers from the four chosen venues} that eventually get cited by patents in our dataset, we compare (a)~the time lag between the publication year of the paper and the issue year of the first patent that cites the research paper (\textit{first patent citation lag}), and (b)~the time lag between the publication year of the paper and the paper's ``peak citation year'' when the research paper gets the most academic citations (\textit{peak citation lag}). 

Peak citation lag averages $5.74$ years in our dataset, compared with $7.48$ years for first patent citation lag.\footnote{The first patent citation lag is lower than patent backward citation lag reported earlier ($10.5$ years) due to right censoring, i.e. recent patent-cited papers are biased towards short lags since those with long lags have not yet been observed in the dataset. Peak citation lag have similar issues. If we allow paper enough time to accrue patent citations, e.g. focus the analysis on papers published before 2000 (cutoff year), we get an average first patent citation lag of 10.4 years (thus replicating the prior results) and peak citation lag of 7.5 years. We varied the cutoff year, and found on average first patent citation lag is always longer than peak citation lag which suggests the robustness of our finding.} A paired t-test confirms that the difference between these two lags are significant $t(3740)=18.3$ $(p<.001)$\rev{, Cohen's D=$0.38$.} This result supports the concern that HCI's focus shifts to other topics by the time industry take up an idea.


\paragraph{\textbf{Self-cite tends to be faster.}}
One exception to this temporal pattern is that self-citation patents have a shorter patent-paper time lag. Since $2008$, the time lag for the non-self-cite patents increased rapidly and was above 14.6 years in $2018$, while the self-cite patents remain below $6.3$ years, which suggests that papers transferred faster by authors themselves into patents compared with those transferred by others. 


\subsection{RQ3: Where is the impact of HCI research on patents?} \label{Paper Content vs. Patent content over time} 
Which HCI research topics are the focus of industry activity? To answer this question, \rev{we compare non-patent-cited HCI papers to patent-cited HCI papers in the four chosen venues via Latent Dirichlet Allocation (LDA)}, a classic method of topic modeling \cite{blei2003latent}. LDA automatically discovers topics within documents, where each topic is represented as a probability distribution of words. Each document can also be represented as a probability distribution over different topics. 

\begin{figure*}[t]
    \centering
    \includegraphics[width=\linewidth]{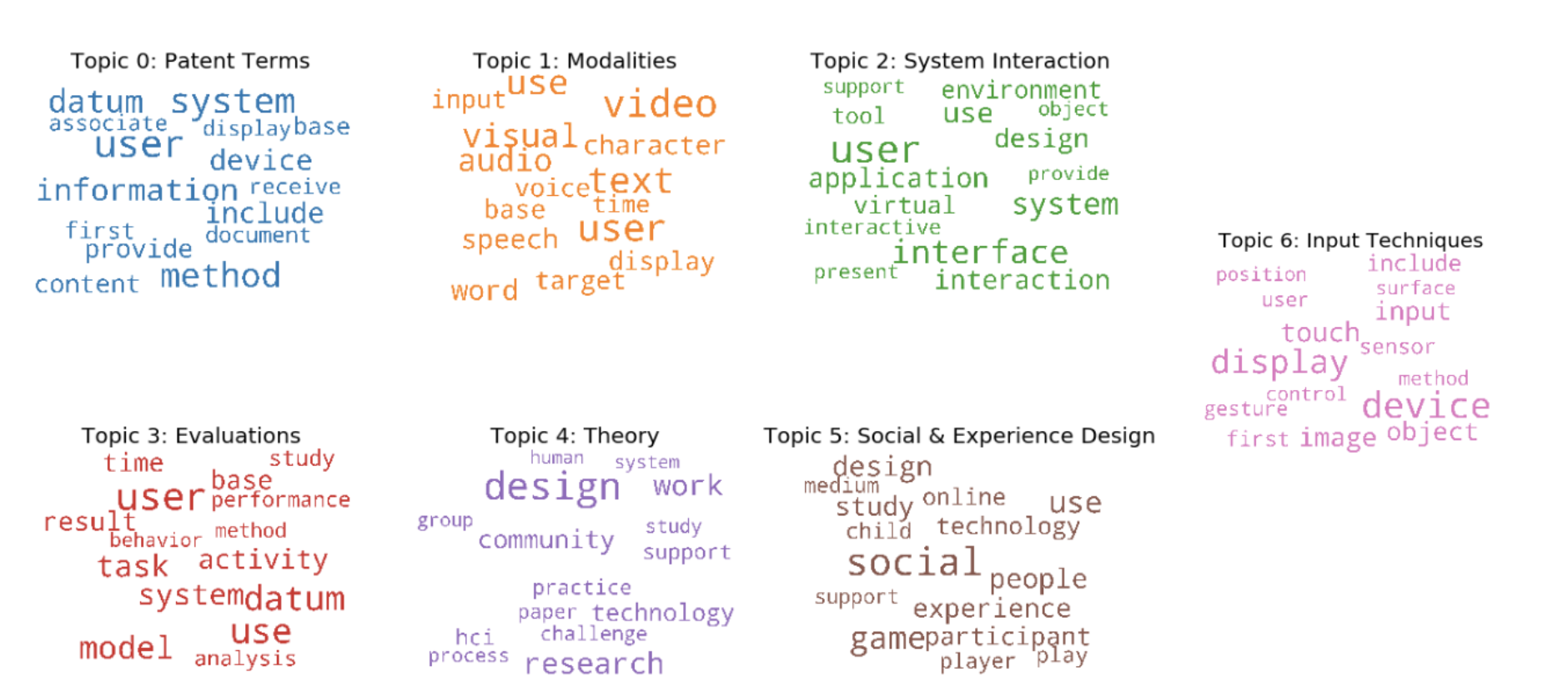}
    \caption{Topics were identified through a Latent Dirichlet Allocation (LDA) analysis of the combined paper-patent corpus. }
    \label{fig:topic_fig2}
\end{figure*}

We concatenated each paper title with its abstract (if available) to represent its contents. Similarly, we concatenated each patent title with its abstract (if available) to represent the patent's contents. We then tokenized the text corpora into unigrams and bigrams, filtered out terms that appear fewer than 5 times in the corpus, removed stop words in English, and then ran LDA modeling. We varied the number of topics and align on seven topics resulting in the highest quality topics.  Figure~\ref{fig:topic_fig2} reports the result. Through checking representative documents and word clusters with HCI experts, we titled each topic: topic $0$ is related to patent terms, the topic is $1$ on modalities, topic $2$ is system interaction, topic $3$ is on evaluations, topic $4$ is on theory, topic $5$ is on social and experience design, and topic $6$ is on input techniques. 

\begin{figure*}[t]
    \centering
    \includegraphics[width=\linewidth]{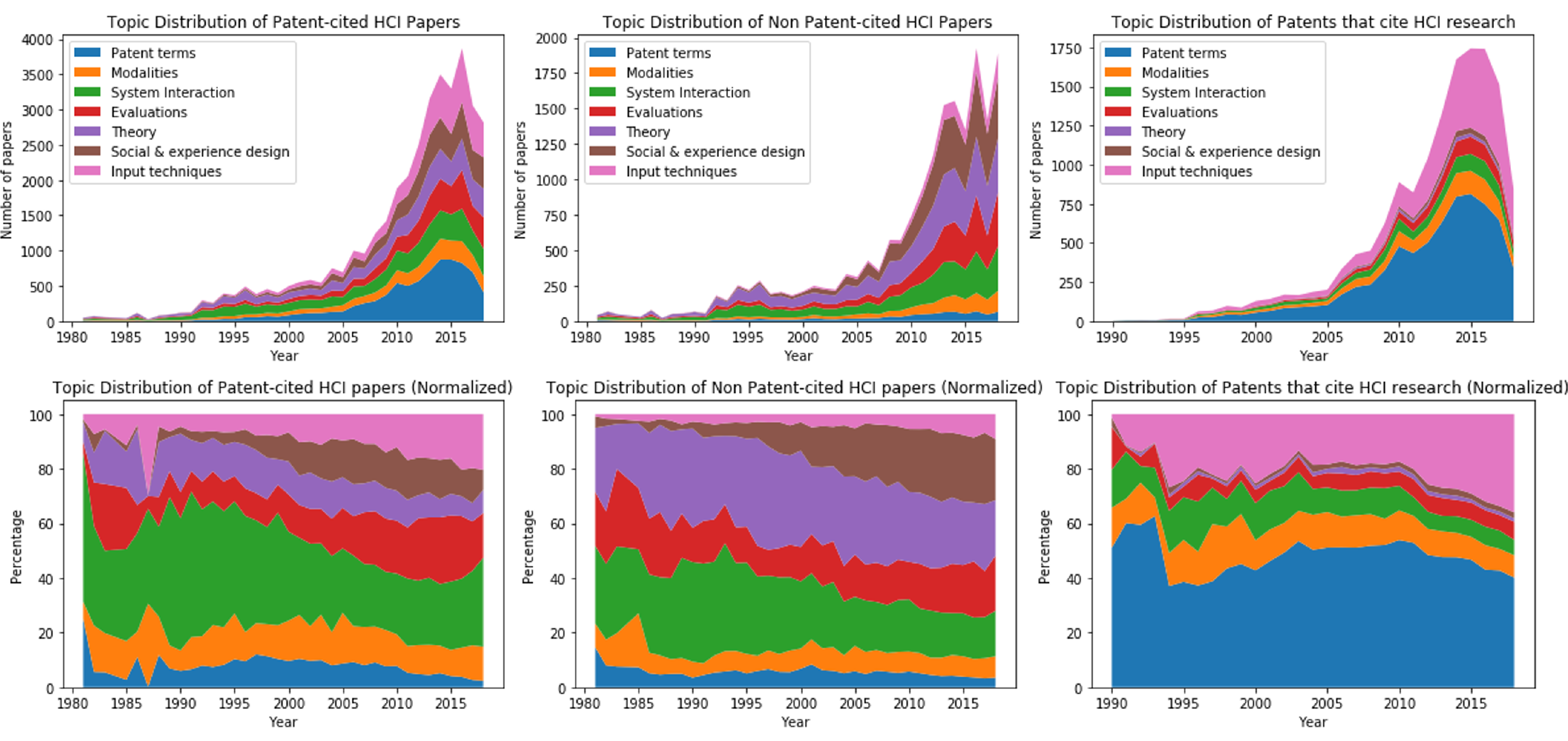}
    \caption{The first row shows the breakdowns of papers across $7$ topics in HCI over time. The second row depicts the percentage of each topic in terms of paper number. Three columns depict "topic distribution of patent-cited HCI papers", "topic distribution of non-patent cited HCI papers" and "topic distribution of patents" respectively.
    System Interaction dominates the patent-cited HCI papers while Theory dominates the non-patent cited HCI papers and Input Techniques dominate patents over time.
    } 
    \label{fig:topic_fig3}
\end{figure*}

We then computed the topic distributions for each document (paper or patent) in our corpus, then aggregated topic distributions of all documents within a specific year that belong to a certain document category (patents, patent cited papers, or non-patent cited papers) so as to get an estimated number of documents that belong to a particular topic for that document category for a particular year. In the first row of Figure~\ref{fig:topic_fig3}, we plotted the topic distribution for patent-cited HCI papers (left), non-patent cited HCI papers (middle), and patents (right), i.e., how many papers belong to topic X in year Y. 
The second row of Figure~\ref{fig:topic_fig3} normalizes this topic distribution, i.e., what is the proportion of topic X in year Y for a specific document category, to better illustrates the distribution pattern.


As can be observed from Figure~\ref{fig:topic_fig3}, system interaction has dominated the patent-cited HCI papers over time, indicating that system-oriented research has been of considerable importance in patent-cited HCI research. From $1980$ to $2000$, about $40$\% patent-cited HCI paper are system interaction related. After $2000$, the percentage of system interaction decreased to about $20$\% but began expanding again in 2015. We also observed that input techniques have expanded significantly over time and reached nearly $20$\% after 2015. Evaluations have also grown in general and contributed about 20\% of all patent-cited HCI papers.

In comparison, the topic distribution of non-patent cited papers shows a very different pattern. The results mirror the methodological plurality of HCI, where not all contribution types have an \rev{industry impact}. Theory work is highly visible in non-patent cited HCI papers over time, though the proportion is gradually decreasing from about 40\% before 2000 to about $20$\% in 2018. Social and experience design has grown significantly from nearly $0$ percent in $1980$ to about $20$\% in $2018$, indicating behavior-oriented research has been of considerable importance in non-patent-cited HCI papers. Evaluations and system interaction contributed to about half of all non-patent-cited HCI papers in 1980, but this percentage has decreased to about 30\% in 2018. Through unpaired t-test, we further verify there exist statistically significant differences between topic distributions in patent-cited papers and non-patent cited papers: there is a higher proportion of theory $(p<.001)$, social \& experience design $(p<.05)$ work, and lower proportion of system interaction $(p<.001)$, modalities $(p<.001)$ work in non-patent-cited HCI papers compared to patent-cited counterparts. \rev{We emphasize that this is not a negative outcome for theory, behavioral, and other research that does not produce artifacts, as they have an impact through other channels, or could influence patent in an indirect way \cite{colusso2019translational}.}

Additionally, the variation of the patents' topic distribution over time is not consistent with that of papers. Since 1990, patent topics have been dominated by input techniques,\footnote{We exclude analysis of topic - `patent terms' as the topic is generic language use in patents.} which first expand from $1990$ to $1993$, then slightly shrink from $1993$ to $2010$ and expand again since $2010$. 
In 2018, about 40\% of patents that cite HCI research papers are input techniques. We also observed this growth in patent-cited HCI papers, but not this significant. 

\subsection{RQ4: Who is involved in the process of recognizing HCI research on patents?}
\rev{Last, we investigate through the four premier HCI venues which institutions are most likely to develop patents that recognize HCI research, and which institutions conduct HCI research that are most cited by patents. Such analysis is important because it identifies the role of different stakeholders within the technology translation landscape~\cite{colusso2019translational}. }


\paragraph{\textbf{Apple, Microsoft, IBM, but no longer Xerox: top \rev{institutes} citing HCI research.}} We examined who are the top patent assignees (the entity that has the property right to the patent, e.g. firm) that cite HCI research. The top patent assignees have been dominated by companies: Apple, Microsoft, and International Business Machines Corporation (IBM) are the top three companies that were granted the highest number of HCI-citing patents in the dataset. Other rise and fall over time. \rev{See appendix C for more details.}

\begin{figure*}[t]
    \centering
    \includegraphics[width=\linewidth]{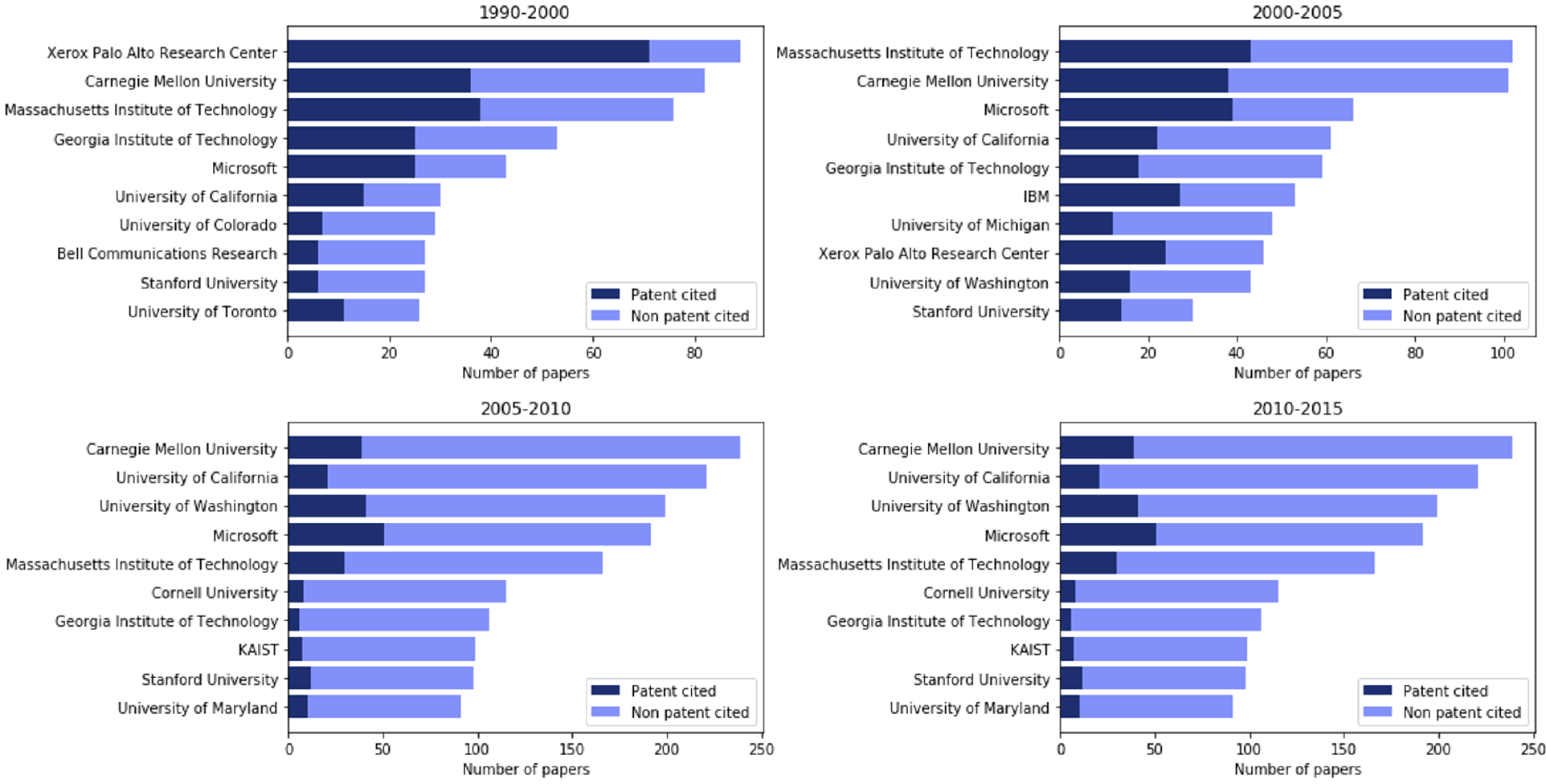}
        \caption{Institutions publishing the most patent-cited research.} 
    \label{fig:f11}
\end{figure*}

\paragraph{\textbf{PARC, CMU, MIT: top institutes that publish patent-cited research.}} We assessed the institutes that published the most patent-cited HCI papers across the years. As Figure~\ref{fig:f11} shows, contrary to the fact that top patent assignees have been dominated by industries, top institutes that published patent-cited HCI papers have been a combination of universities and companies. Top universities include Carnegie Mellon University, Massachusetts Institute of Technology, University of California, and University of Washington. Top companies that published patent-cited HCI papers include Xerox Palo Alto Research Center and Microsoft. The ratio of patents cited among all HCI papers significantly dropped from nearly half before 2005 to less than $30$\% for most institutes after $2005$, due to the fact that the total number of HCI papers grew significantly and the right censoring issue. 



Overall, $35.5\%$ of Microsoft’s papers, $31.0\%$ of IBM’s, and $65.1\%$ of Xerox’s were cited by patents. In comparison, universities have a lower rate of papers cited by patents, e.g. 25.2\%, 15.3\%, 26.9\% of papers were recognized among Carnegie Mellon University, the University of California system, and MIT respectively. This indicates that among institutes publishing the most HCI papers, the papers from the industry have a higher proportion of papers recognized by patents. However, the difference between industry and universities becomes smaller when removing self-citing patents.

\paragraph{\textbf{Self-citation.}} We also explored the degree of self-citation. We find that 13.9\% of patents self-cite the inventor's own research. Although the number of self-citing patents is growing, the percentage of self-citations in all HCI patent citations is decreasing from around 20\% to 5\% in recent years. This suggests that while the HCI field is expanding, the number of researchers directly referring to their own research in patents is not growing at the same rate. Most of the self-citations also come from industry, with Microsoft and Xerox constituting 34.8\% and 11.2\% of total self-citations. Self-citation from academia is much less common.  





\textbf{Summary of conclusions:} Through our analysis, we find that HCI research has had a significant impact on patents, with an increasing number of patents recognizing research in CHI, CSCW, UIST, and UbiComp. Patents are more likely to refer to systems-oriented and highly-cited research in academia. However, the time lag between patent and paper is long (>10 years) and getting longer, suggesting HCI research and practice may be inefficiently connected. We further verify the robustness of our main findings through two additional analyses, which we report in Appendix D.

\section{Discussion}
In this section, we discuss the implications of our findings: 
 
\subsection{\rev{The patent-research relevance landscape in HCI}}

By combining the findings from our large-scale analyses with that of prior qualitative evidence established by literature (e.g. case studies~\cite{chilana2015technology}, personal experience,~\cite{czerwinski2003hci} and interviews~\cite{colusso2019translational}), we can now offer a more comprehensive picture of the HCI translation landscape.


\rev{\paragraph{\textbf{The impact of HCI research on patents:}} Our work largely corroborates literature arguing for the considerable impact of HCI research on practice~\cite{harrison2018hci, buxton2008long, myers2000past}. In our analysis, among HCI research papers in CHI, CSCW, UIST, and UbiComp, 20.1\% of all papers have been referenced by patents, and 13.4\% for SIGCHI sponsored venues overall. This is a rate far higher than science in general (1.5\%~\cite{marx2020reliance}) and prominent journals across multiple scientific fields (9.7\% \cite{bryan2020text}). The rate is also higher than bio-medicine, a field that has a more systematic technology translation system and a richer tradition of studying technology translation, whose proportion is 7.7\% ~\cite{manjunath2021comprehensive}.\footnote{Bio-medicine papers from US institutes only---a filter we did not apply for our study of HCI---have a proportion of 23.3\%~\cite{manjunath2021comprehensive}, which is roughly the same as HCI research.}
HCI research diffuses into the industry at a similar rate as Computer Vision (25\%) and at a higher rate than NLP (11\%), both areas of substantial industry funding and interest.}
Note our estimate is a lower bound: given the long time lag of patent-paper citations, recent papers may have not fully expressed their impact yet (right censoring). When only considering earlier years that do not suffer much from right censoring issues (e.g. prior to 2005), we see roughly 30\%-50\% of papers published in those years have been cited by patents. For UIST, the proportion is even higher, close to 80\% for many years. 



\paragraph{\textbf{\rev{Issues with the current HCI translation into patents:}}} As argued by Bill Buxton in `the long nose of innovation'~\cite{buxton2008long}, the bulk of innovation takes place over a long period: the mouse was first built in 1965 by William English and Doug Engelbart, but was only popularized in the 1990s when Microsoft released a large-scale commercial mouse; multitouch was published in 1985, but took 22 years to become a product. \rev{Our analysis further demonstrates that even the initial step of having research recognized in a patent, which may be well before there is an actual product, takes considerable time.} In fact, the ubiquity of long time delay between research and practice, and thus lack of immediate impact on the industry after the publication of a research paper, could be one underlying reason why many papers on HCI translation argue that HCI lacks practical impact~\cite{colusso2017translational,czerwinski2003hci,rynes2012research}. Furthermore, our analysis demonstrates that the time lag between patent and research is getting longer over time, indicating that the translation process in HCI may become more inefficient over time.
This result is in line with a general trend across science (average over time: 14.4 years), where they report an average patent citation to science time lag of about 8 years in the 1990s, rising to about 15 years in 2018~\cite{marx2022reliance}. The specific reason for the (increasing) time lag would need further work. 
We also show that the HCI community often leaves an idea behind by the time industry gets interested, as a paper's peak citation lag is generally shorter than the paper's first patent citation lag. The result indicates that with a long time lag, HCI research has moved on and is exploring new emerging technologies that are not yet reliable enough, cheap enough, power-efficient enough, or accurate enough for the industry yet. The observation supports the observation that HCI research often plays ``the time machine game'',\footnote{A term attributed to Jeff Pierce, formerly a research manager at IBM Research and faculty member at Georgia Tech.} where it fast forwards into the future by acquiring early versions of emerging technology (e.g., VR, AR, multi-touch, AI) and exploring the interactive applications of that technology. Unless HCI is directly working on reducing those barriers to industry entry for that technology, HCI research cannot directly accelerate the time lag: it is simply painting a compelling vision of the future before that future arrives.


\subsection{\rev{How could the HCI community do better to facilitate technology transfer and industrial impact?}}
\paragraph{\textbf{Encourage communications and collaborations across academia and industry.}} 
Through our analysis, we have found that even though research articles from both academia and industry are recognized by patents, the proportion of papers in academia recognized by patents is much lower. While the result could be that industry research papers by themselves are more applied than research papers from academia, or that industry has more internal incentives to have their research patented\footnote{Microsoft Research, for example, would award decorative ``patent cubes'' to researchers for each new patent they co-authored, which researchers would often stack into decorative pyramids and display in their offices}, this could also be a sign that practitioners are not fully aware of some application-oriented advances in academia, and that information diffusion between academia and practice is inefficient~\cite{buxton2008long}. 


Our work thus echoes calls for a more inclusive and translation-friendly environment~\cite{chilana2015technology, colusso2017translational, colusso2019translational, buie2010bring}: that both academia and industry should 1)~better recognize the importance of technology translation rather than considering translation irrelevant, 2)~establish more communication and collaboration channels to engage people, e.g. SIGGRAPH-style Emerging Tech festivals where academic researchers show their published HCI work to an applied audience and encourage researchers in serving as advising role in the industry, and 3)~involve more HCI materials in Computer Science curriculum at universities to get `future practitioners' more familiar with HCI research ideas, and thus prepare them as translational developers who are more likely to bridge academia and industry~\cite{norman2010research} 

\paragraph{\textbf{Encourage self-driven technology transfer.}}
Self-driven technology transfer (e.g. patents recognizing one's own paper) generally happens much faster than technology transfer in general. Intuitively, the self-driven transfer would not encounter many of the same communication and information diffusion barriers. Self-driven technology transfer could also potentially solve many of the `recognition' issues in the translational process as discussed in prior works~\cite{harrison2018hci}. However, as shown in our analysis, though the amount of self-driven technology transfer in HCI is going up over time, it is not on par with the rate of increase for research articles. While not all researchers should actively engage in technology transfer, there could be more steps to be taken to encourage self-driven technology transfer from the academic side so that translation could happen more efficiently, e.g. through better supporting and recognizing attempts to self-translate one's own research by providing legal apparatuses and funding support. Meanwhile, we want to emphasize while there are benefits of self-driven transfer, it may currently not distribute opportunities equally. For instance, in the life sciences~\cite{ding2006gender}, women faculty members patent at about 40\% of the rate of men. It would be important to identify and mitigate these potential issues so as to ensure an inclusive technology transfer environment. \rev{Relatedly, as suggested by prior work \cite{colusso2019translational}, there exist multiple translational gaps in HCI, and basic researchers should also be encouraged to engage more with applied researchers and do more system work, which would eventually help translate HCI research insights into industry impacts.}


\paragraph{\textbf{Recognizing translational work in HCI}}
More broadly, our work echoes prior work on the need of recognizing translational efforts in HCI. For instance, when allocating funding or considering researcher promotion, their impacts in the industry could be taken into consideration as a separate metric aside from impacts within academia. \rev{Our work points to a potential way to quantify one important pathway towards HCI research's impacts on the industry, through analyzing patent-to-science citation data.}

\paragraph{\textbf{Impact signals}}

Prior approaches to quantifying research impact mostly focus on impact within academia through bibliometric analysis. \rev{However, no quantitative metric fully captures the complexities of our world. Could the h-index be fruitfully complemented with other information? (a ``patent relevance'' p-index?) While our analysis show impacts in academia and impacts in patents correlate, we also find papers with high patent citations do not necessarily have high paper citations: in one extreme case, the most patent-cited paper in our dataset, ``A multi-touch three-dimensional touch-sensitive tablet''~\cite{lee1985multi}, is more popular in the patent world than in academia. If evaluations primarily consider the academic impacts of such research work, the work's value may have been underestimated. As one potential pathway to industry impacts that are relatively easy to scale, patents provide a potential signal to more holistically evaluate research.}

\rev{Of course, patent relevance, or practice relevance in general\footnote{\rev{Though arguably it's much harder to quantify other forms of practice relevance, e.g. how research influence design patterns and open source software}}, is not the solitary metric of scientific value, and research and researchers should not be judged based on a single metric, e.g. to receive funding or get a promotion.} Thus, our work should not be interpreted as stating that non-patent cited research represents any sort of failure. There are many, many examples of influential HCI research that is not patented (or even patentable). For instance, our work shows that system building or application-oriented HCI research is more likely to find relevance in patents rather than design-oriented or behavioral research. \rev{The result is not an indication that applied-oriented research is more valuable: there could be the indirect influence of other types of works on application-oriented research, e.g. applied research getting inspiration from behavior work, as suggested by the translational science model in HCI \cite{colusso2019translational} -- which we seek to address in future work, and 2) it is equally important to maintain a diversity of research ideas, which has proven to facilitate greater innovation for science in general~\cite{hofstra2020diversity}. If the measurement of this impact is desirable, we will require new methods, such as multi-hop influence over citation network \cite{ahmadpoor2017dual}, linguistic concept diffusion \cite{cao2020will}, from the paper to the public or media~\cite{yin2022public, yin2021coevolution}.}





\subsection{Limitations and Future Work} \label{limitation}
\rev{Patent citations to research are only a proxy signal of industry impact, which is a hard-to-quantify concept otherwise. It is only one, among many (e.g. open source software, design patterns), potential pathway to industry impacts. First, not all patents will turn into products or practices, so they may not be actual ``industry impact'' instances (false positives). There could be many other factors, such as assignee strategy and resources, that could influence the process. Even if a patent does end up as a product, most of the time the patent will not be valuable or impactful, with 97\% of all patents never recouping the cost of filing them\footnote{\url{https://www.forbes.com/sites/stephenkey/2017/11/13/in-todays-market-do-patents-even-matter/}}. However, the fact that inventors decide to go through the long and expensive process of filing a patent to protect their intellectual property does indicate they are considering their invention having at least some potential to be of relevance to the practice domain, which could be regarded as an intended act aiming at industry impact or technology transfer.}

Second, industry impact could happen even if there is no patenting process involved (false negative), which is not uncommon in software~\cite{graham2005software}: startups will launch products without patents from time to time, which is quite different from the innovation landscape of more traditional fields; \rev{design processes (e.g., usability testing, heuristic evaluation), design patterns, and open source software (e.g., d3, Vega Lite) also have significant industry impact that is not reflected though patents.} As such, our analysis of using patent citation to HCI research papers could be different from the actual translation landscape: the patent dataset could introduce both false positives and negatives, e.g., even if a patent cites a HCI paper, it may never be taken up in practice as product, and an actual product that gets influenced by HCI research that is unpatentable will not be observed and measured through our current approach.

Despite all the shortcomings of patent citation to science, the availability and scale of the dataset make it a rare lens in the innovation literature to enable conclusions on the research-practice relationship at scale~\cite{manjunath2021comprehensive,marx2020reliance, marx2022reliance}. In our work, in addition to building on these methods from the innovation literature, we tied our analysis to qualitative evidence discussed in prior works so as to validate our findings. 

In future work, we plan to 1) involve more qualitative evidence (e.g. interviewing inventors' motivation behind citing HCI research) to further validate our findings, and 2) take more steps to quantify how HCI research turns into \textit{valuable} inventions, e.g. by using patent citations to other patents as a proxy of patent value, which correlates well with other metrics of patent value, e.g. whether they are renewed to a full term, and whether they get licensed~\cite{harhoff1999citation, sampat2004patent}.



\rev{Our work also currently mostly focuses on measuring industry relevance at the paper level}, which may not necessarily be the principal unit of knowledge: for example, several papers on the same idea can get cited by patents. While we have made preliminary attempts to analyze the topics prevalent to patents, patent-cited research papers, and non-patent cited research papers, future work could better study at the concept level what specific research ideas are transferred into research, either through keywords provided by the author (which is unfortunately not available in our current dataset), or natural language processing based approach such as phrase mining~\cite{cao2020will}, which may help track transfer of innovations at a more fine-grained level.

Other limitations include: \rev{(1)~our dataset is focused on United States patents, which limits our cultural context and generalizability, though arguably a significant proportion of inventors/organizations using (and pushing) HCI research in practice are US-based \cite{sturm2015weird};} (2)\rev{~ while discussing in a descriptive way in our paper with findings on the role of academic impacts (section 4.1), topic (section 4.3), and institute/actors (section 4.4) in relating to patent impact, we do not have causal evidence/analysis on the causal mechanisms what cause some papers to have more industry relevance, which is an important topic we seek to address in future work}, and (3)~if there are recent trends in the last 5-10 years that have changed these patterns, it is still too recent to see their impact.



\section{Conclusions}

In this work, drawing inspiration from the innovation literature, \rev{we quantitatively study one important pathway from HCI research to industry impact by conducting a large-scale analysis of how patent documents from USPTO refer to research articles in CHI, CSCW, UIST, UbiComp and other SIGCHI sponsored venues. We contribute to the literature by measuring to what extent HCI research has been featured in patent citations, with a high proportion of papers referenced in patents.} Patents are more likely to refer to systems-oriented and highly-cited research in HCI. However, we also reveal potential translation issues: HCI research and practice may not be efficiently coupled, since the time lag between paper and patent is long and getting longer. \rev{Our work not only demonstrates the potential of using patent citation data to science as a powerful tool to study the industry impact of HCI research, but also points to suggestions for the HCI community to better facilitate translation from research to practice.}

\begin{acks}
The authors thank Yian Yin for helpful suggestions on polishing the work, and Mary Czerwinski, Bongshin Lee, Lucy Lu Wang, James Zou, Shumin Zhai and many others for insightful discussions. Hancheng Cao was supported by Stanford Interdisciplinary Graduate Fellowship.
\end{acks}

\bibliographystyle{ACM-Reference-Format}
\bibliography{reference}

\newpage
\appendix
\section{details of data acquisition}
\label{adix:data_detail}
Here we provide details of the data acquisition procedure that generate our final analyzed data.

\paragraph{\textbf{Patent citation to science that connects USPTO to Microsoft Academic Graph}}
To capture the information required by patent citation to science, we utilize a public dataset available over Zenodo.\footnote{\url{http://relianceonscience.org}} We leverage the patent‐to‐article citations of Version v$37$ (Jul 19, 2022), including \textit{\_pcs\_mag\_doi\_pmid.tsv} and \textit{papercitations.tsv}.
For \textit{\_pcs\_mag\_doi\_pmid.tsv}, we mainly focus on the fields \texttt{reftype}, \texttt{diff\_month}, \texttt{selfciteconf\_avg}.
 We focus on fields \texttt{citingpaperid} and \texttt{citedpaperid} in \textit{papercitations.tsv}, which we used to join with Microsoft Academic Graph Metadata.

\paragraph{\textbf{Microsoft Academic Graph Metadata}}
Microsoft Academic Graph Metadata is also available over Zenodo.\footnote{\url{http://relianceonscience.org}} The data files we utilize include \textit{authoridname\_normalized.tsv}, \textit{conferenceidname.tsv}, \textit{paperauthoridaffiliationname.tsv}, \textit{paperauthororder.tsv}, \textit{paperconferenceid.tsv} and \textit{paperyear.tsv}.

\paragraph{\textbf{USPTO Metadata}}
We acquire USPTO metadata from PatentsView.\footnote{\url{https://patentsview.org/download/data-download-tables}} We utilize datafiles \textit{assignee}, \textit{inventor}, \textit{patent}, \textit{patent\_assignee}, and \textit{patent\_inventor}.

\paragraph{\textbf{Semantic Scholar}}

We request Semantic Scholar API\footnote{\url{https://api.semanticscholar.org/api-docs/graph##tag/Paper-Data/operation/get_graph_get_paper_references}} with research article IDs retrieved from Microsoft Academic Graph Metadata for extra paper information. The fields we queried include \texttt{title}, \texttt{abstract}, \texttt{venue}, \texttt{year}, \texttt{referenceCount}, \texttt{citationCount}, \texttt{authors}, as well as \texttt{name}, \texttt{affiliations}, \texttt{paperCount}, and \\
\texttt{citationCount} associated with each author.

we retrieved all the above data in Aug 2022.


\rev{\section{SIGCHI Sponsored Venues}} \label{adix:sigchi}
The 20 SIGCHI venues that we include in our analysis are: Human Factors in Computing Systems (CHI), User Interface Software and Technology (UIST), Ubiquitous Computing (UbiComp), Conference on Computer Supported Cooperative Work (CSCW), Conference on Tangible and Embedded Interaction (TEI), Symposium on Eye Tracking Research \& Application (ETRA), International Conference on Supporting Group Work (GROUP), Conference on Intelligent User Interfaces (IUI), Creativity and Cognition (C\&C), Interaction Design and Children (IDC), International Conference on User Modeling, Adaptation, and Personalization (UMAP), Symposium on Engineering Interactive Computing System (EICS), Conference on Automotive User Interfaces and Interactive Vehicular Applications (AutomotiveUI), Conference on Human-Robot Interaction (HRI), International Conference on Computational Collective Intelligence (CI), Conference on Recommender Systems (RecSys), Annual Symposium on Computer-Human Interaction in Play (CHI PLAY), International Conference on Multimodal Interaction (ICMI), Symposium on Spatial User Interaction (SUI), Symposium on Virtual Reality Software and Technology (VRST). 

\rev{In total, there are 57,385 papers where 13.4\% of them (7678 papers) have been cited by patents in our dataset.}

\begin{figure*}[t]
    \centering
    \includegraphics[width=\linewidth]{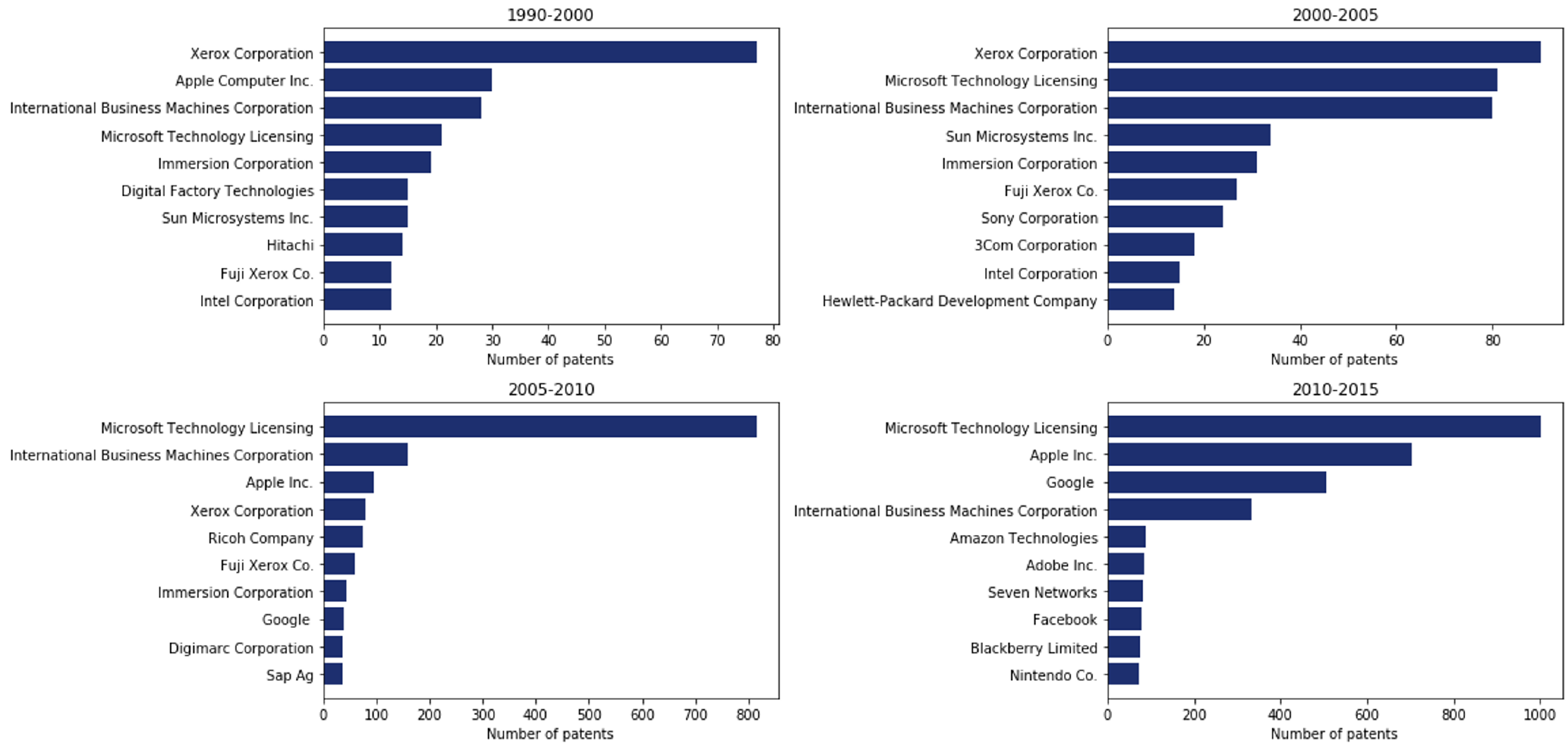}
        \caption{\rev{Top patent assignees that cite HCI research over time.} }
    \label{fig:f7}
\end{figure*}

\rev{\section{Top patent assignees over time}}
\rev{We show top patent assignees over time in Fig \ref{fig:f7}.}



\rev{\section{Additional Analysis on non-self-citing patents and non-researcher patents}}
\label{adix:add_analysis}
We provide two additional analyses using a subset of \rev{four premier venues} to further verify the robustness of our findings. To rule out the possibility that the impacts of HCI research on patents is a result of self-cite, or driven primarily by HCI researchers -- thus one may argue the impact of HCI research in industry is actually limited --  we run the same analysis using  1) patents that do not include self-cite to one's own research papers (``non-self-citing patents'')), which is $26,382$ ($86.04\%$ of original patents), and 2) patents that are invented by people who have never published any CHI, CSCW, UIST or UbiComp research papers ( (``non-researcher'' patents), which we operationalized through excluding patents where inventor last name have appeared in author lists of papers from the four venues we focused on.\footnote{This set of patents is a smaller set than actual  ``non-researcher'' patents. The primary objective is to ensure a set of patents with inventors who, for sure, have never published papers in the four academic venues we studied without tedious author disambiguation.} This results in $5,251$ ($17.12\%$ of original patents) of ``non-researcher'' patents. We find consistent patterns in our main analysis where a high proportion of HCI research papers are cited by patents, and there is a long time lag between patent and paper. More specific results are as follows:



\begin{figure*}[t]
    \centering
    \includegraphics[width=\linewidth]{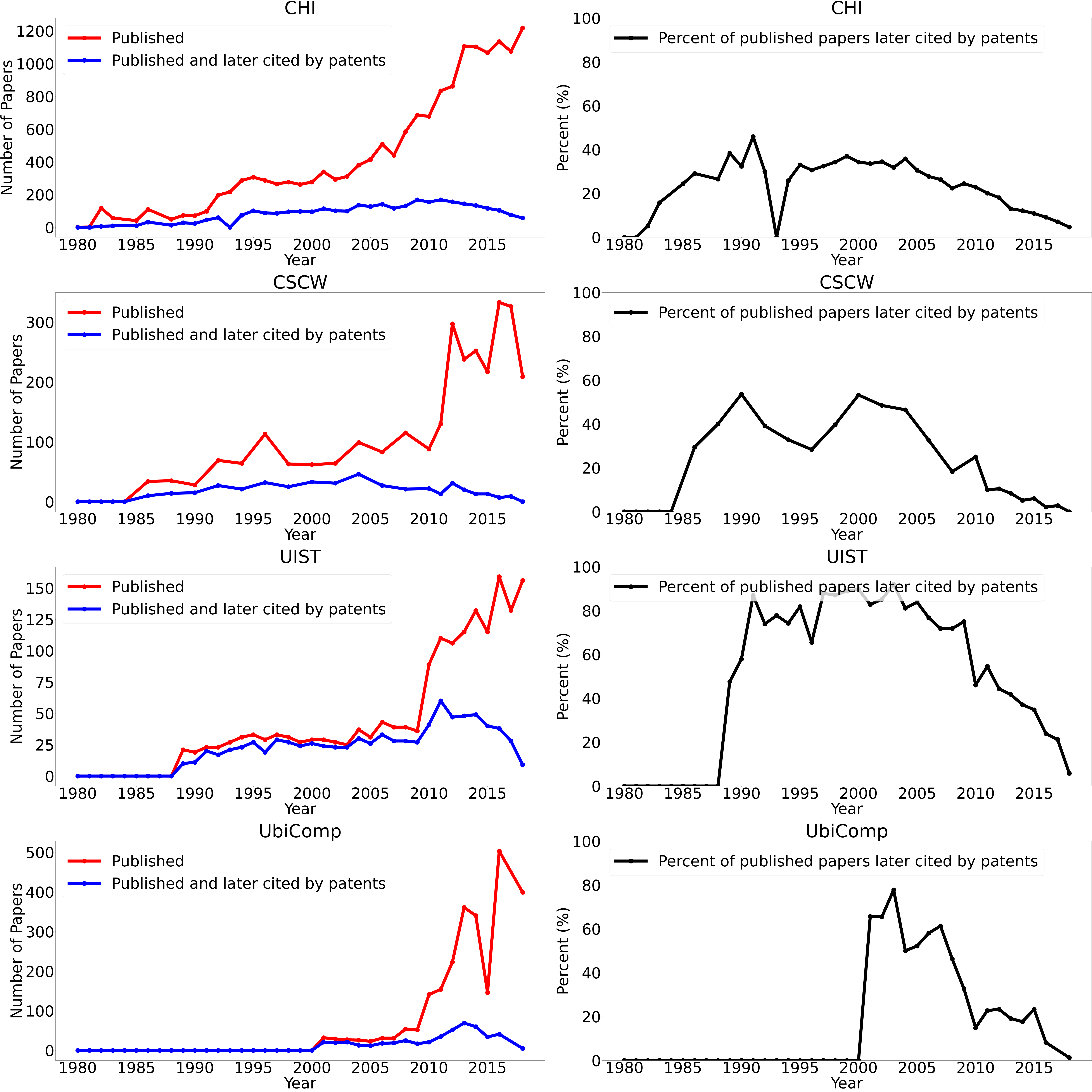}
    \caption{(Non-self-cite) Left: the number of papers published by each conference per year (red) and the number of papers published in that year that were later cited by at least one patent (blue), at ACM CHI, CSCW, UbiComp, and UIST.
    } 
    \label{fig:b_f1_nr}
\end{figure*}

\begin{figure*}[t]
    \centering
    \includegraphics[width=0.5\linewidth]{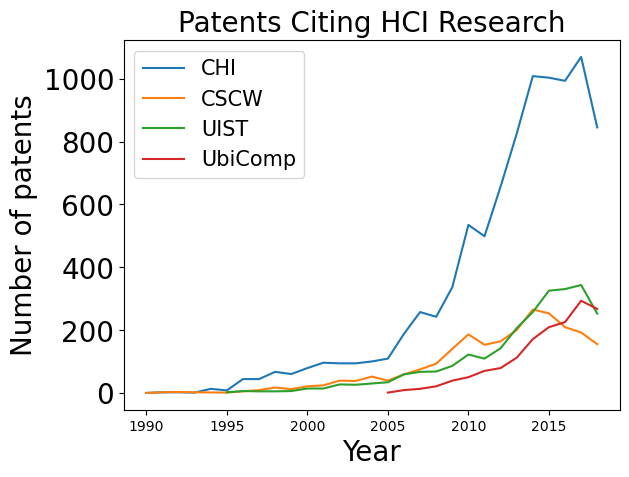}
    \caption{(Non-self-cite) The number of patents that cite HCI papers over time.}
    \label{fig:b_f2_nr}
\end{figure*}


\begin{figure*}[t]
    \centering
    \includegraphics[width=\linewidth]{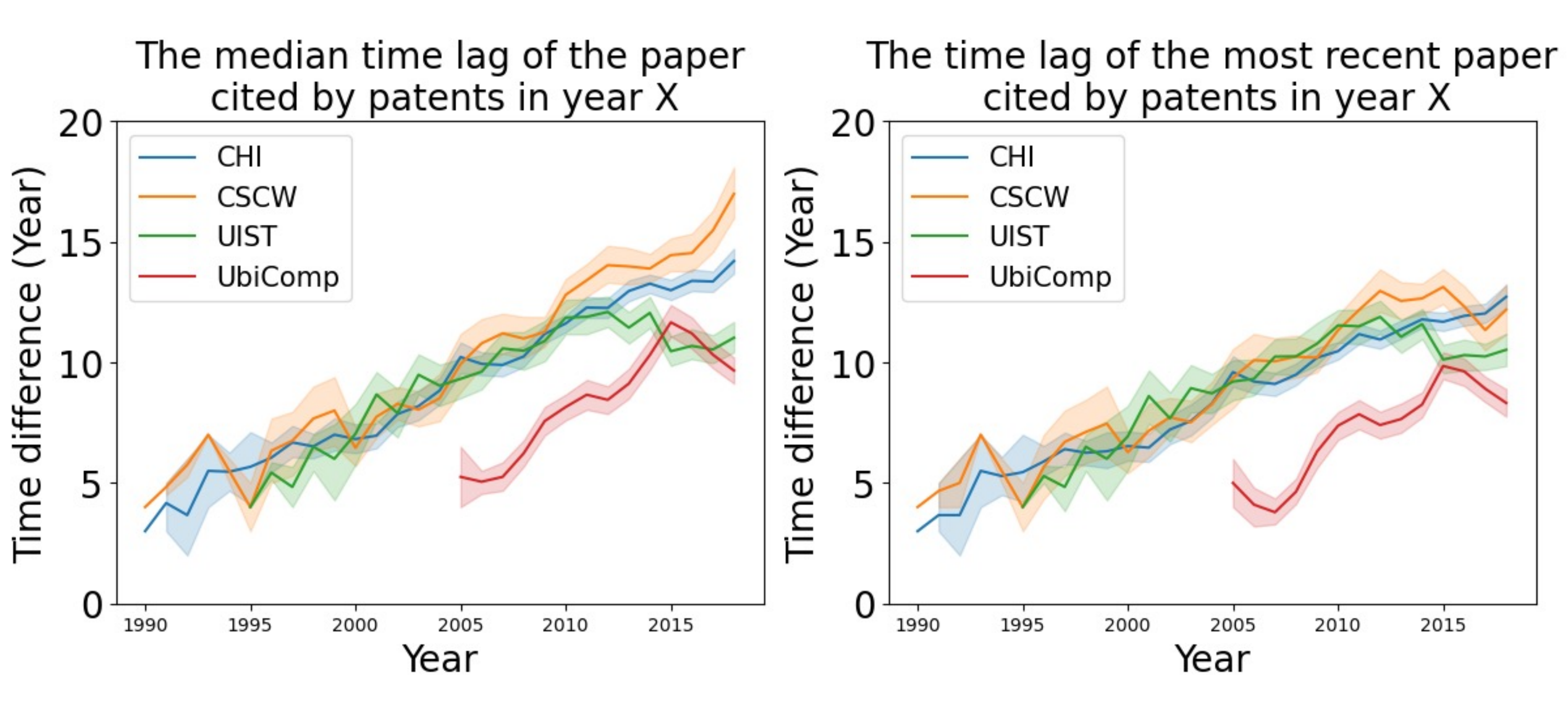}
        \caption{(Non-self-cite) The time lag between patent and paper is long and getting longer for different types of citations and venues.} 
    \label{fig:a_f9}
\end{figure*}

\paragraph{\textbf{Proportion of papers that get cited by patents.}}
The proportion of papers cited by non-self-citing patents is plotted in Figure ~\ref{fig:b_f1_nr} and the ratio rises and persists since 1990 at over 30\%. At UIST in particular, the patent citation ratio reaches $60\%$ - $80\%$ from $1990$ - $2010$. This suggests that non-self-citing patents, similar to our main result, recognize a considerable number of HCI research papers.

\begin{figure*}[t]
    \centering
    \includegraphics[width=\linewidth]{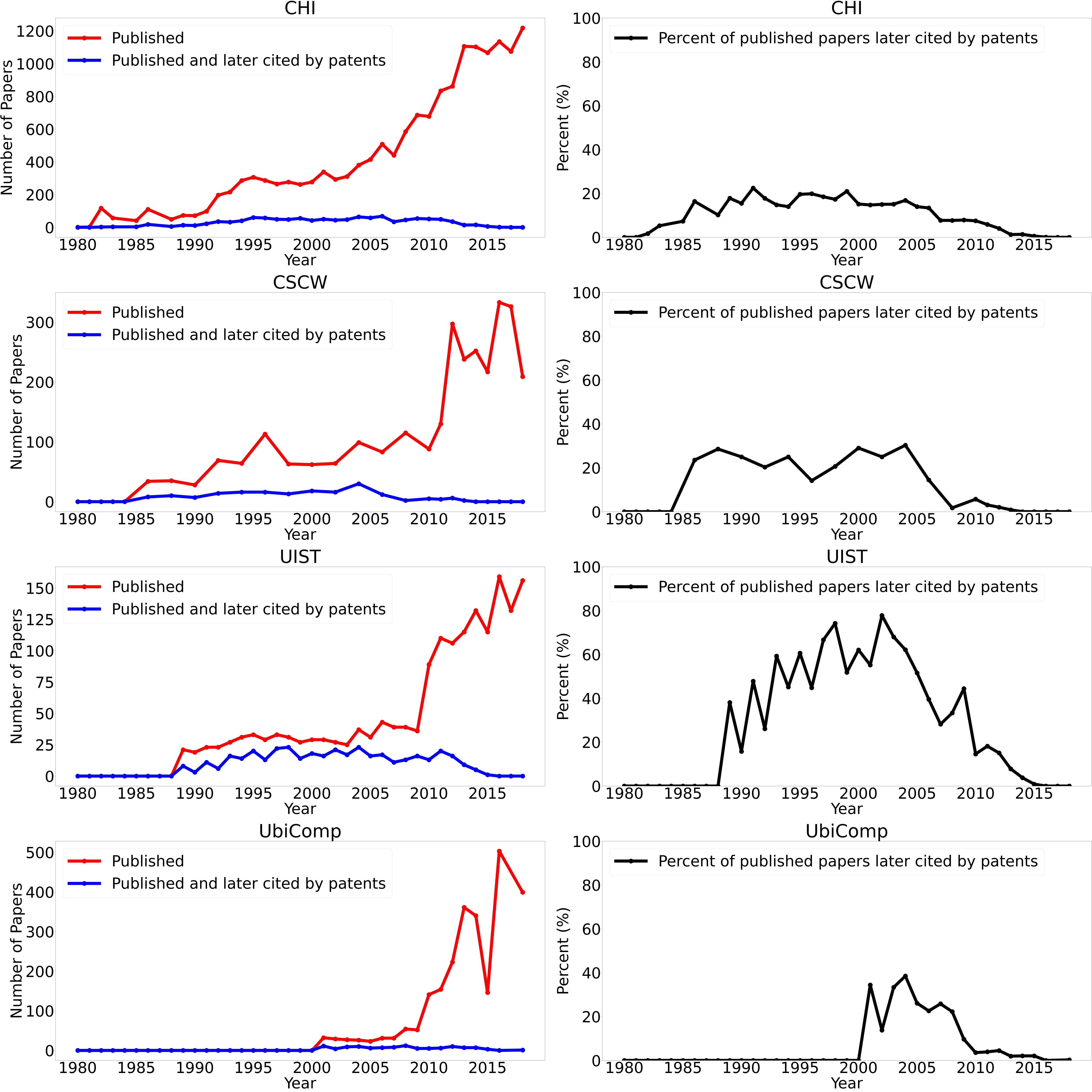}
    \caption{(Non-researcher) Left: the number of papers published by each conference per year (red) and the number of papers published in that year that were later cited by at least one patent (blue), at ACM CHI, CSCW, UbiComp, and UIST.
    } 
    \label{fig:a_f1_nr}
\end{figure*}

Identical trends can be observed for non-researcher patents, as shown in Figure ~\ref{fig:a_f1_nr}.

\paragraph{\textbf{Increasing citations to HCI research in patents.}}
Figure~\ref{fig:b_f2_nr} shows the number of non-self-citing patents that cite HCI research over time. It can be observed that non-self-citing patents first increase  in 2000 and then peak around 2014, ranging from $200$ to $1000$ across different venues. This agrees with the overall trend reported in the main paper.

\begin{figure*}[t]
    \centering
    \includegraphics[width=0.5\linewidth]{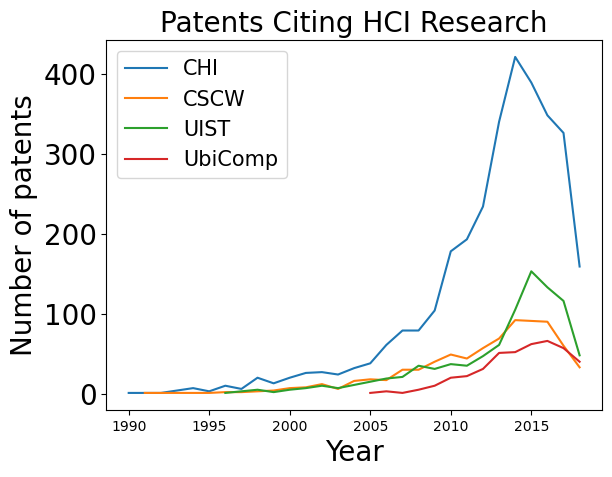}
    \caption{(Non-researcher) The number of patents that cite HCI papers over time.}
    \label{fig:a_f2_nr}
\end{figure*}

Identical trends can be observed for non-researcher patents, as shown in Figure ~\ref{fig:a_f2_nr}.

\begin{figure*}[t]
    \centering
    \includegraphics[width=\linewidth]{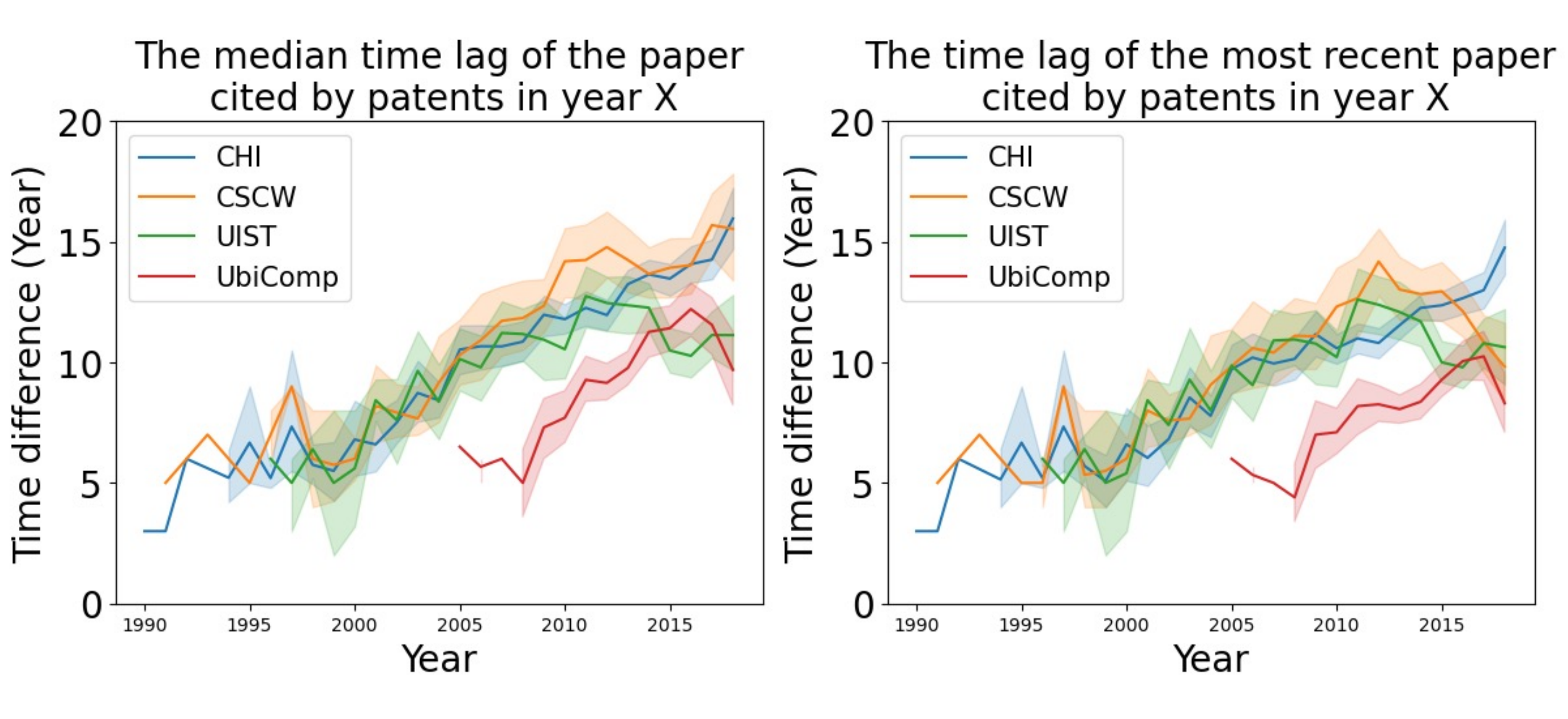}
        \caption{(Non-researcher) The time lag between patent and paper is long and getting longer for different types of citations and venues.}
    \label{fig:b_f9}
\end{figure*}

\paragraph{\textbf{Time lag between patent and paper is long and getting longer.}}
The temporal trend of the measured time lag between the issue date of non-self-citing patents and the publication date of HCI papers they cited are plotted in Figure ~\ref{fig:b_f9}a. Similar to the trend reported in the main results (Figure ~\ref{fig:f9}), the median time lag increased from 1989 to 2014 for all the venues from about around $5$ years to around $10-15$ years while since 2014, this trend bifurcates among different venues. The time lag between the patent and its most recent cited paper (Figure  ~\ref{fig:b_f9}b ) is also examined, showing identical trends.

Identical trends can be observed for non-researcher patents, as shown in Figure ~\ref{fig:a_f9}.

\end{document}